\documentclass{amsart} 
\usepackage{graphicx}
\usepackage{amssymb, amsmath, sidecap}
\usepackage{amsfonts}
\usepackage{amssymb,hyperref}
\usepackage{float}
\usepackage{tikz} 
\usetikzlibrary{arrows,shapes,positioning,shadows,trees}

\newcommand{\C}{\mathbb C}

\newcommand{\R}{\mathbb R}

\def\la{\label}

\def\op{\otimes_p}
\def\qfor{\qquad \text{ for }}

\def\bpr{\begin{pri}}
\def\epr{\end{pri}}

\def\eld{\frac{d}{d\lambda}\Big|_{\lambda=0}}

\def\be{\begin{equation}}
\def\ee{\end{equation}}
\def\br{\begin{rem}}
\def\er{\end{rem}}
\def\bi{\begin{itemize}}
\def\ei{\end{itemize}}
\def\tilde{\widetilde}

\def\bt{\begin{thm}}
\def\et{\end{thm}}
\def\bl{\begin{lem}}
\def\el{\end{lem}}
\def\bd{\begin{defi}}
\def\ed{\end{defi}}
\def\bc{\begin{cor}}
\def\ec{\end{cor}}
\def\bp{\begin{proof}}
\def\ep{\end{proof}}
\def\br{\begin{rem}}
\def\er{\end{rem}}

\newtheorem{thm}{Theorem}[section]
\newtheorem{lem}{Lemma}[section]
\newtheorem{defi}{Definition}[section]

\newtheorem{rem}{Remark}[section]
\newtheorem{cor}{Corollary}[section]

\newtheorem{pri}[thm]{Principle}
\newtheorem{interaction}[thm]{Geometric Interaction Mechanism}
\newtheorem{y-interaction}[thm]{Yukawa Interaction Mechanism}

\newtheorem{DIM}[thm]{Duality of Interaction Mediators}
\newtheorem{DIF}[thm]{Duality of Interaction Forces}

\numberwithin{equation}{section}
\numberwithin{figure}{section}

\begin{document}
\title[Unified Field Equations Coupling Four Forces]{Unified Field Equations Coupling Four Forces  and Principle of Interaction Dynamics}
\author[Ma]{Tian Ma}
\address[TM]{Department of Mathematics, Sichuan University,
Chengdu, P. R. China}

\author[Wang]{Shouhong Wang}
\address[SW]{Department of Mathematics,
Indiana University, Bloomington, IN 47405}
\email{showang@indiana.edu, http://www.indiana.edu/~fluid}

\thanks{The authors are grateful for the referee's insightful comments and suggestions. The work was supported in part by the
Office of Naval Research, by the US National Science Foundation, and by the Chinese National Science Foundation.}


\begin{abstract} The  main objective of this article is to postulate a principle of interaction dynamics (PID) and to derive unified field equations coupling the four fundamental interactions based on first principles.  PID is a least action principle subject to div$_A$-free constraints for the variational element  with $A$  being gauge potentials.  The Lagrangian action is uniquely determined by 1)  the principle of general relativity, 2) the $U(1)$, $SU(2)$  and $SU(3)$ gauge invariances, 3) the Lorentz invariance, and 4) principle of principle of representation invariance (PRI), introduced in \cite{field2}. The unified field equations are then derived using PID. The unified field model spontaneously  breaks the gauge symmetries, and   gives rise to a new mechanism for energy  and mass generation. 
The unified field model introduces a natural duality between the mediators and their dual mediators, and can be easily decoupled to study each  individual interaction when other interactions are negligible. The unified field model, together with PRI and PID applied to individual interactions, provides clear explanations and solutions to a number of outstanding challenges in physics and cosmology, including e.g. the dark energy and dark matter phenomena, the quark confinement, asymptotic freedom, short-range nature of both strong and weak interactions, decay mechanism of sub-atomic particles, baryon asymmetry, and the solar neutrino problem.
\end{abstract}
\maketitle
\tableofcontents

\section{Introduction}
The four fundamental forces/interactions of Nature are the electromagnetic interaction, the strong interaction, the weak interaction and the gravity. 
Current successful  theories describing these
interactions include the Einstein general theory of
relativity for  gravitation,  
  and the Standard Model, built upon an $SU(3)\otimes SU(2)\otimes
U(1)$ gauge theory for electromagnetic, the weak and the strong interactions; see among many others \cite{kaku, quigg, griffiths, kane, halzen}. Apparently a unified field theory will be built upon the success of gauge theory  and the Einstein general theory of relativity. 
There are, however, a number of unexplained mysteries of Nature, including for example the dark energy and dark matter phenomena, quark confinement, asymptotic freedom, short-range nature of both strong and weak interactions, decay mechanism of sub-atomic particles, and the solar neutrino problem.

The main objectives of this article are 1) to postulate a new fundamental principle, which we call the principle of interaction dynamics (PID), and 2) to derive a unified field theory coupling  all four interactions based on the following fundamental first principles:
\begin{itemize}
\item the principle of general relativity, 
\item the principle of gauge invariance,
 \item the principle of Lorentz invariance, 
\item the principle of interaction dynamics (PID), and
\item the principle of representation invariance (PRI).

\end{itemize}

PID is introduced in this article, and PRI is postulated in \cite{field2}. Intuitively, PID takes the variation of the action functional under energy-momentum conservation constraint. PRI requires that physical laws be independent of representations of the gauge groups.

Hereafter we describe the main ideas and ingredients of this article. 

\medskip

\noindent
{\it Dark matter and dark energy as evidence for PID} 

\medskip

There are strong physical evidences for the validity of PID. The first is the discovery of dark matter and dark energy. As the law of gravity, the Einstein gravitational field equations are inevitably needed to be modified to account for dark energy and dark matter. Over the years, there are numerous  attempts, which can be classified into two groups: 1) $f(R)$ theories, and 2) scalar field theories. Also, the chief current attempts for dark matter are both direct and indirect searches for  dark matter particles such as the weakly interactive massive particles (WIMPs).  Unfortunately,  both dark energy and  dark matter are still two greatest mysteries in modern physics. 

In \cite{MW12}, we attack this problem in the fundamental level {\it based on first principles}. Since Albert Einstein discovered the general theory of relativity in 1915, his two fundamental  first principles, the principle of equivalence (PE) and the principle of general relativity (PGR) have gained strong and decisive observational supports. These two principles amount to saying that the space-time manifold is a 4-dimensional (4D) Riemannian manifold $M$ with the Riemannian metric being regarded as the gravitational potentials. 

The law of gravity is then represented by the gravitational field equations solving for the gravitational potentials $\{g_{\mu\nu}\}$, the Riemannina metric of the space-time manifold $M$.  The principle of general relativity is a fundamental symmetry of Nature. This symmetry principle, together with the simplicity of laws of Nature, uniquely determines the Lagrangian action for gravity  as the Einstein-Hilbert functional: 
\be\la{eh-action}
L_{EH}(\{g_{\mu\nu}\})  = \int_M \left(R+ \frac{8\pi G}{c^4} S\right)\sqrt{-g}dx.
\ee
Here $R$ stands for the scalar curvature of $M$, and $S$ is the energy-momentum density of matter field  in the universe. 
In fact, this very symmetry is the main reason why the current $f(R)$ and scalar field theories can only yield certain approximations of the law of gravity.

We observe in \cite{MW12} that due to the presence of dark energy and dark matter, the energy-momentum tensor $T_{\mu\nu}$ of normal matter  is in general  no longer conserved:
$$
\nabla^\mu(T_{\mu\nu})\not= 0.
$$

By the Orthogonal Decomposition Theorem \ref{t2.17}, the Euler-Lagrange variation of the Einstein-Hilbert functional $L_{EH}$ is uniquely balanced by the co-variant gradient of a vector field $\Phi_\mu$: $\nabla_\mu \Phi_\nu$, leading to
the following new  gravitational field equations:
\begin{equation}
\begin{aligned}
& 
R_{\mu\nu}-\frac{1}{2}g_{\mu\nu}R=-\frac{8\pi
G}{c^4}T_{\mu\nu} + \nabla_{\mu}\Phi_{\nu},\\
& \nabla^\mu\left[ \frac{8\pi
G}{c^4}T_{\mu\nu} - \nabla_{\mu}\Phi_{\nu} \right] =0.
\end{aligned}
\label{gfe}
\end{equation}
which give rise to a unified theory for dark matter and dark energy \cite{MW12}. 

Equivalently, as we have shown in \cite{MW12} that the Euler-Lagrangian variation of the Einstein-Hilbert functional $L_{EH}$ must be taken  under  energy-momentum conservation constraints; see  \cite{MW12} for details:
\begin{align}
\frac{d}{d\lambda}\Big|_{\lambda = 0} L_{EH}(g_{\mu\nu}+ \lambda X_{\mu\nu})  
=  &   (\delta L_{EH}(g_{\mu\nu}), X)  \la{cel} \\
= &  0\quad \forall X=\{X_{\mu\nu}\}  \text{ with } 
\nabla^\mu X_{\mu\nu}=0. \nonumber 
\end{align}

The term $\nabla_\mu\Phi_\nu$ does not correspond to any Lagrangian action density, and is the direct consequence of energy-momentum conservation constraint of the variation  element $X$ in (\ref{cel}).  For the case given here, the vector field 
$\Phi_\nu$  is in fact the gradient of a scalar field $\varphi$: 
$\Phi_\nu =- \nabla_\nu \varphi$. However, if we  take the cosmic microwave background radiation into consideration,  the field equations are in a more general form with the vector field $\Phi_\nu$; see  (\ref{(3.4.5)}):
\begin{equation}
R_{\mu\nu}-\frac{1}{2}g_{\mu\nu}R=-\frac{8\pi
G}{c^4}T_{\mu\nu}+(\nabla_{\mu}+\frac{e}{\hbar
c}A_{\mu})\Phi_{\nu},\label{cmb-gfe}
\end{equation}
where the term $\frac{e}{\hbar c}A_{\mu}\Phi_{\nu}$ represents the
coupling between the gravitation and the microwave background
radiation.

We have shown that it is the duality  between  the {\it attracting} gravitational field $\{
g_{\mu\nu}\}$  and the {\it repulsive} dual field $\{\Psi_\mu\}$, and their nonlinear interaction that give rise to gravity, and in particular the gravitational effect of dark energy and dark matter; see \cite{MW12} and Section \ref{s5.2}.

In a nutshell, the gravitational field equations (\ref{gfe}) or (\ref{cmb-gfe}) are derived based on the principle of equivalence and the principle of general relativity, which uniquely dictate the specific form of the Einstein-Hilbert action. 
The energy-momentum conservation constraint variation (\ref{cel}) is simply the {\it direct and unique consequence} of the presence of dark energy and dark matter. Hence it is natural for us to postulate PID for all four fundamental interactions, which amounts to variation of  the Lagrangian action under the div$_A$-free constraints, where $A$ represent the gauge potentials.

\medskip

\noindent{\it Symmetries of fundamental interactions}

\medskip
Fundamental  laws  of Nature  are  universal, and their validity is independent of the space-time location and directions of  experiments and observations. 
The universality of laws of Nature  implies
that  the Lagrange actions are invariant  and the differential equations are covariant under certain symmetry. 

As discussed early,  the law of gravity is determined by the principle of general relativity. The electromagnetic, the weak and the strong interactions are dictated, respectively,  by the  $U(1)$, the $SU(2)$  and $SU(3)$ gauge invariances. In other words,  following the simplicity principle of laws of Nature, the three basic  symmetries---the Einstein general relativity, the
Lorentz invariance and the gauge invariance---uniquely determine 
the interaction fields and their Lagrangian actions for the four
interactions. For example, the $SU(2)$  gauge invariance   for the weak interaction uniquely determines the Lagrangian density:
\begin{align}
& \mathcal{L}_W=-\frac{1}{4}G^w_{ab}W^a_{\mu\nu}W^{\mu\nu b}, \la{weak-density}
\end{align}
which represents the scalar curvature of the complex vector 
bundle\footnote{Throughout this article, we use the notation $\op$ to denote "gluing a vector space to each point of a manifold" to form a vector bundle. For example, 
$$M\op \C^n=\cup_{p \in M} \{ p \} \times \C^n$$
is a vector bundle with base manifold $M$ and fiber vector space $\C^n$.},  
$M\op (\C^4)^2$,   of Dirac spinor fields. 
Here for $a=1, 2, 3$, 
\begin{align*}
& W^a_{\mu}=(W^a_0,W^a_1,W^a_2,W^a_3), &&
W^a_{\mu\nu}=\partial_{\mu}W^a_{\nu}-\partial_{\nu}W^a_{\mu}+g_w\lambda^a_{bc}W^b_{\mu}W^c_{\nu}
\end{align*}
are weak gauge potentials and the corresponding curvature associated with the connection on $M\op (\C^4)^2$:
$$D_{\mu} =\nabla_{\mu}+ig_wW^a_{\mu}\sigma_a,$$
where $U =e^{i\theta^a(x) \sigma_a} \in SU(2)$,   $\{\sigma_a \ | \ a=1, 2, 3\}$ is a set of generators for $SU(2)$,  $g_w$  is the gauge coupling constant, $\lambda^a_{bc}$ are the  structure constants of  $\{\sigma_a\}$, and  
$$
G^w_{ab}=\frac{1}{8} \lambda^c_{ad}\lambda^d_{cb} =\frac12 \text{Tr}(\tau_a \tau_b^\dagger).
$$

The Lagrangian density (\ref{weak-density}) obeys also PRI, which was discovered and postulated by the authors  in \cite{field2}. In other words, the physical quantities 
$W^a_\mu$, $W^a_{\mu\nu}$,  $\lambda^a_{bc}$  and $G^w_{ab}$ are $SU(2)$-tensors under the following transformation of representation generators:
\begin{equation}
\tilde{\sigma}_a=x^b_a\sigma_b,\label{g-transf}
\end{equation}
where $X=(x^b_a)$ is a nondegenerate complex matrix.

One profound consequence of PRI is that any linear combination of gauge potentials from two different gauge groups are prohibited by PRI. For example, the term $\alpha A_\mu + \beta W^3_\mu$ in the electroweak theory violates PRI, as  this term does not represent a gauge potential for any gauge group. In fact, the term combines one component of a tensor with another component of an entirely different tensor with respect to the transformations of  representation generators as given by  (\ref{g-transf}). 

\medskip

\noindent
{\it PID as a mechanism for gauge symmetry breaking and mass generation} 

\medskip

The principle of general relativity, the Lorentz invariance  and  PRI stand for the universality of physical laws, i.e., the validity of
laws of Nature is independent of the coordinate systems expressing them. Consequently,  these symmetries hold true for both the Lagrangian actions and their  variational equations. 

The physical implication of the gauge symmetry is different. Namely, the gauge symmetry holds true only for the Lagrangian actions for the electromagnetic,
week and strong interactions, and it will be broken in the field
equations of these interactions. This is a general principle, which we call the principle of gauge symmetry breaking. 

The principle of gauge symmetry breaking can be regarded as part of the spontaneous symmetry breaking, which is a phenomenon
appearing in various physical fields. Although the phenomenon was discovered in superconductivity by
Ginzburg-Landau in 1951, the mechanism of spontaneous 
symmetry breaking in particle physics was first proposed by Y. Nambu in 1960; see \cite{nambu60, nambu-jona1, nambu-jona2}. This mechanism amounts to saying
that a physical system, whose underlying laws are invariant
under a symmetry transformation, may spontaneously break the
symmetry, if this system possesses some states that don't satisfy
this symmetry.

The Higgs mechanism is a special case of the Nambo-Jona-Lasinio
spontaneous symmetry breaking, a gauge symmetry
breaking mechanism, leading to the mass  generation  of sub-atomic particles. 
This mechanism was discovered 
at almost the same time in 1964  by three groups of six physicists  \cite{higgs, englert, guralnik}.

PID discovered in this article provides a new principle  to achieve  gauge
symmetry breaking and  mass generation. The difference between
both PID and Higgs mechanisms  is that the first one is a natural
sequence from the first principle, and the second one is to add
artificially a Higgs field in the Lagrangian action. In addition,
the PID obeys PRI, and the Higgs  violates PRI.  

\medskip

\noindent
{\it Interaction mechanism}

\medskip

One of greatest revolutions  in sciences is Albert Einstein's vision on gravity: the gravitational force is caused by the space-time curvature.  Yukawa's viewpoint, entirely different from Einstein's, is that the other three fundamental forces take place through exchanging intermediate bosons such as photons for the electromagnetic interaction, W$^\pm$ and Z intermediate vector bosons for the weak interaction, and gluons for the strong interaction. 

Based on the unified field theory presented in this article and in \cite{field2}, in the same spirit as the Einstein's principle of equivalence of gravitational force, it is natural for us to  postulate an alternate mechanism for all four interactions:  The gravitational force is the
curved effect of the time-space, and the electromagnetic, weak,
strong interactions are the twisted effects of the underlying complex vector 
bundles $M\op \C^n$.

\medskip

\noindent
{\it Unified field theory}

\medskip

We have demonstrated that the Lagrangian actions for fundamental interactions are uniquely determined by 
\begin{itemize}

\item the principle of general relativity,

\item the principle of gauge invariance,

\item the principle of representation invariance (PRI), and 

\item principle of Lorentz invariance. 

\end{itemize}

Based on PRI, the coupled  Lagrangian action for the four fundamental interactions is naturally given by 
 \begin{equation}
L=\int_M\left[\mathcal{L}_{EH}+\mathcal{L}_{EM}+\mathcal{L}_W+\mathcal{L}_S+ \mathcal{L}_D+\mathcal{L}_{KG}\right]\sqrt{-g}dx\label{4-lagrangian}
\end{equation}
where  $\mathcal{L}_{EH}$, $\mathcal{L}_{EM}$, $\mathcal{L}_W$  and $\mathcal{L}_S$  are the Lagrangian densities for the gravity, the electromagnetism, the weak and the strong interactions,  and $\mathcal{L}_D$  and $\mathcal{L}_{KG}$  are  the actions for both Dirac spinor fields and Klein-Gordon fields.

This action is invariant under the following symmetries: general relativity, the Lorentz invariance, PRI, and the gauge invariance as given in Section \ref{s3.3}.

With the Lagrangian action $L$   at our disposal, the unified field equations are then derived using PID; see equations (\ref{(3.3.9)})-(\ref{(3.3.15)}). 
These equations are naturally covariant under the symmetries: general relativity, the Lorentz invariance and PRI. As indicated before, the unified field equations  spontaneously break the $U(1)$, $SU(2)$ and $SU(3)$ gauge symmetries, giving rise to a new mass generation mechanism, entirely different from the Higgs mechanism. 

\bigskip

\noindent
{\it Duality of fundamental interactions} 

\medskip
In the unified field equations (\ref{(3.3.19)})-(\ref{(3.3.22)}), 
 there exists a natural duality between the
interaction fields $(g_{\mu\nu},A_{\mu},W^a_{\mu},S^k_{\mu})$ and
their corresponding dual fields $(\phi^G_{\mu},\phi^E,\phi^w_a,\phi^s_k)$ :
\be\label{duality}
\begin{aligned} 
& g_{\mu\nu}  &&  \leftrightarrow && \phi^G_{\mu}, \\
&A_{\mu}  &&  \leftrightarrow  &&  \phi^E,\\
&W^a_{\mu} && \leftrightarrow   && \phi^a_w && \text{for } 1\leq a\leq 3, \\
&S^k_{\mu} && \leftrightarrow  &&  \phi^k_s  && \text{for } 1\leq k\leq 8.
\end{aligned}
\ee
The duality  relation (\ref{duality}) can be regarded as a duality  
between field particles for each interaction. It is clear  that each interaction 
mediator possesses  a dual field particle, called the dual mediator,
and if the mediator has spin-$k$, then its dual mediator has spin-$(k-1)$.
Hence the dual field particles consist of spin-1 dual graviton, spin-0 dual photon, spin-0 charged Higgs  and neutral Higgs fields, and spin-0 dual gluons. 
The neutral Higgs $H^0$ (the dual particle of $Z$)
had been discovered experimentally. 

In the weakton model \cite{weakton}, we realize that these dual particles possess the same weakton constituents, but different spins, as the mediators.

Thanks to  the PRI symmetry, the $SU(2)$ gauge fields
$W^a_{\mu}\ (1\leq a\leq 3)$ and the $SU(3)$ gauge fields
$S^k_{\mu}\ (1\leq k\leq 8)$ are symmetric in their indices
$a=1,2,3$ and $k=1,\cdots ,8$ respectively. Therefore, the
corresponding relation (\ref{duality}) can be considered as a duality of interacting forces:  Each interaction
generates both attracting and repelling forces. Moreover, for the corresponding  pair
of dual fields, the even-spin field generates an attracting force,  and  the
odd-spin field generates a repelling force.
 
For the first time, we discovered  such attracting and repelling property of each 
interaction derived from the field model. Such property plays crucial role in the stability of matter in the Universe. For example,  repulsive  behavior of gravity on a very large scale we discovered in \cite{MW12} explains dark energy phenomena. Also, the strong interaction potentials in (\ref{strong-layered}) below demonstrate, for example, that  as the distance between  two  quarks increases, the strong force is  repelling,  diminishes (asymptotic freedom region), and then becomes attracting (confinement). 

 \bigskip
 
 \noindent
 {\it Decoupling}
 
 \medskip
The unified field model can be easily decoupled to study each  individual interaction when other interactions are negligible. In other words, PID is certainly applicable to each individual interaction. For gravity, for example, PID offers  a new gravitational field model,  leading to a unified model for dark energy and dark matter \cite{MW12}.

\bigskip
 
 \noindent
 {\it Interaction potentials  and force formulas}
 
 \medskip
With PRI and  the duality for both  strong and weak interactions,  we are able to derive the long overdue strong and weak  potentials and force formulas.  

In fact, we have derived in \cite{strong}  the  layered formulas of strong interaction potentials for various level particles. In particular,  the
$w^*$-weakton potential $\Phi_0$, the quark potential $\Phi_q$, the
nucleon/hadron potential $\Phi_n$ and the atom/molecule potential
$\Phi_a$  are given as follows \cite{strong}:
\be\label{strong-layered}
\begin{aligned}
&\Phi_0=g_s\left[\frac{1}{r}-\frac{A_0}{\rho_w}(1+k_0r)e^{-k_0r}\right], && \frac{1}{k_0}=10^{-18}\text{ cm},\\
&\Phi_q=\left(\frac{\rho_w}{\rho_q}\right)^3g_s\left[\frac{1}{r}-\frac{A_q}{\rho_q}(1+k_1r)e^{-k_1r}\right], && \frac{1}{k_1}=10^{-16}\text{
cm},\\
&\Phi_n=3\left(\frac{\rho_w}{\rho_n}\right)^3g_s\left[\frac{1}{r}-\frac{A_n}{\rho_n}(1+k_nr)e^{-k_nr}\right], && \frac{1}{k_n}=10^{-13}\text{ cm},\\
&\Phi_a=N\left(\frac{\rho_w}{\rho_a}\right)^3g_s\left[\frac{1}{r}-\frac{A_a}{\rho_a}(1+k_ar)e^{-k_ar}\right], &&\frac{1}{k_a}=10^{-10}\sim 10^{-7}\text{ cm}.
\end{aligned}
\ee

In each of these potentials, the first part $1/r$ gives repelling force and is mainly due to gluon fields, and the second part leads to attracting force and is due to the dual fields. This clearly demonstrates the need of the dual fields in the strong interaction as required by the strong force confinement property. 

Also with PRI and the duality of weak interaction, we have derived in \cite{weak}   the following layered weak interaction potential formulas:
\begin{equation}
\begin{aligned}
& \Phi_w=g_w(\rho)e^{-kr}\left[\frac{1}{r}-\frac{B}{\rho}(1+2kr)e^{-kr}\right], &&  
\frac{1}{k}=10^{-16}\text{ cm}, \\
& g_w(\rho )=N\left(\frac{\rho_w}{\rho}\right)^3g_w,
\end{aligned}\label{weak-p}
\end{equation}
where $\Phi_w$ is the weak force potential of a particle with radius
$\rho$ and $N$ weak charges $g_w$ of each weakton \cite{weakton},  $\rho_w$ is the weakton radius,   and $B$ is a parameter depending on the particles.

\bigskip

\noindent
{\it Quark confinement, asymptotic freedom, and short-range nature of weak and strong interactions}

\medskip

The above weak and strong  potentials  offer a clear mechanisms for quark confinement, for asymptotic freedom,  and for short-range nature of both weak and strong interactions;  see e.g \cite{strong, field2} for details. 

\bigskip
\noindent{\it Weakton model, baryon asymmetry, and solar neutrino problem}

Thanks also to these potentials, we are able to derive in \cite{weakton} a weakton model  of elementary particles, leading to an explanation of all known sub-atomic decays and the creation/annihilation  of matter/antimatter particles, as well as the baryon asymmetry problem. 

Remarkably,  in the weakton model,   both  the spin-1 mediators (the photon, the W and Z vector bosons, and the gluons) and the spin-0 dual mediators introduced in the unified field model in this article have the {\it same} weakton constituents, differing  only by their spin arrangements.  The spin arrangements  clearly demonstrate that  there must be dual mediators with spin-0. This observation clearly provides another strong evidence of PID and the unified field model introduced in this article. 

Also, the weakton model offer an alternate explanation to the longstanding solar neutrino problem.  When the solar electron neutrinos collide with anti-electron neutrinos, which are abundant due to the $\beta$-decay of neutrons, they can form  
$\nu$ mediators and flying away, causing the loss of electron neutrinos.
This is consistent with the experimental results on the agreement between the speed of light and the speed of neutrinos. 

\section{Principle of Interaction Dynamics (PID)}

\subsection{PID}
The main objective in this  section is to propose  a fundamental
principle of physics, which we call the principle of interaction dynamics (PID). Intuitively, PID takes the variation of the action functional under energy-momentum conservation constraint. 

There are strong physical evidence and motivations for the validity of PID, including 
\begin{enumerate}
\item  the discovery of dark matter and dark energy, 
\item the non-existence of solutions for the classical Einstein
gravitational field equations in general cases,
\item the principle of spontaneous gauge-symmetry breaking, and 
\item  the theory of Ginzburg-Landau superconductivity.
\end{enumerate}

It is remarkable that the term $\nabla_{\mu}\Phi_{\nu}$ in (\ref{gfe}) plays a similar role as the Higgs field in the standard model in particle physics, and the constraint Lagrangian dynamics gives rise to a new first principle, which we call the principle of interaction dynamics (PID). This new first principle provides an entirely different approach to introduce the Higgs field, and leads  to a new mass generation mechanism. 


Let $(M, g_{\mu\nu})$ be the 4-dimensional space-time Riemannian
manifold with $\{g_{\mu\nu}\}$ the Minkowski type Riemannian metric.
For an $(r,s)$-tensor $u$ we define the $A$-gradient and
$A$-divergence operators $\nabla_A$ and $\text{ div}_A$ as
\begin{eqnarray*}
&&\nabla_A u =\nabla u+u\otimes A,\\
&&\text{div}_Au=\text{ div}\ u-A\cdot u,
\end{eqnarray*}
where $A$ is a vector field and here stands for a gauge field,
$\nabla$ and div are the usual gradient and divergent covariant
differential operators. Let $F=F(u)$ be a functional of a tensor
field $u$. A tensor $u_0$ is called an extremum point of $F$ with
the $\text{ div}_A$-free constraint, if $u_0$ satisfies the equation
\begin{equation}
\frac{d}{d\lambda}\Big|_{\lambda=0} F(u_0+\lambda X) =\int_M\delta
F(u_0)\cdot X\sqrt{-g}dx=0\qquad  \forall X \text{ with } \text{ 
div}_AX=0.\label{(3.1.35)}
\end{equation}

\bpr[Principle of Interaction Dynamics] 
\begin{enumerate}

\item For all
physical interactions there are Lagrangian actions
\begin{equation}
L(g,A,\psi )=\int_M\mathcal{L}(g_{\mu\nu},A,\psi
)\sqrt{-g}dx,\label{(3.1.36)}
\end{equation}
where $g=\{g_{\mu\nu}\}$ is the Riemannian metric representing the
gravitational potential, $A$ is a set of vector fields representing
the gauge potentials, and $\psi$ are the wave functions of
particles. 

\item The action (\ref{(3.1.36)}) satisfy the invariance of
general relativity, Lorentz invariance, gauge invariance and the
gauge representation invariance. 

\item The states $(g,A,\psi )$
are the extremum points of (\ref{(3.1.36)}) with the $\text{
div}_A$-free constraint (\ref{(3.1.35)}).

\end{enumerate}

\epr

Based on PID and Theorems \ref{t2.25}  and \ref{t2.26}, the field equations with
respect to the action (\ref{(3.1.36)}) are given in the form
\begin{align}
&\frac{\delta}{\delta g_{\mu\nu}}L(g,A,\psi
)=(\nabla_{\mu}+\alpha_bA^b_{\mu})\Phi_{\nu},\label{(3.1.37)}\\
&\frac{\delta}{\delta A^a_{\mu}}L(g,A,\psi
)=(\nabla_{\mu}+\beta^a_bA^b_{\mu})\varphi^a,\label{(3.1.38)}\\
&\frac{\delta}{\delta\psi}L(g,A,\psi )=0\label{(3.1.39)}
\end{align}
where $A^a_{\mu}=(A^a_0,A^a_1,A^a_2,A^a_3)$ are the gauge vector
fields for the electromagnetic, the weak and strong interactions,
$\Phi_{\nu}=(\Phi_0,\Phi_1,\Phi_2,\Phi_3)$ in (\ref{(3.1.37)}) is a
vector field induced by gravitational interaction, $\varphi^a$ is the
scalar  fields generated from the gauge field $A^a_{\mu}$, and
$\alpha_b, \beta^a_b$ are coupling parameters.

PID is based on variations with $\text{ div}_A$-free constraint
defined by (\ref{(3.1.35)}). Physically, the 
conditions $$\text{ div}_AX=0\ \ \ \ \text{ in\ (\ref{(3.1.35)})}$$
stand for the  energy-momentum conservation constraints.

\subsection{Principle of representation invariance (PRI)}
We end this section by recalling the principle of representation invariance (PRI) first postulated in \cite{field2}.  We proceed with  the
$SU(N)$ representation. In a neighborhood $U\subset SU(N)$ of the unit
matrix, a matrix $\Omega\in U$ can be written as
$$\Omega =e^{i\theta^a\tau_a},$$
where
\begin{equation}
\tau_a=\{\tau_1,\cdots ,\tau_K\}\subset T_eSU(N),\ \ \ \
K=N^2-1,\label{(3.1.40)}
\end{equation}
is a basis of generators  of the tangent space $T_eSU(N)$. An $SU(N)$
representation transformation is a linear transformation of the basis in 
(\ref{(3.1.40)}) as
\begin{equation}
\tilde{\tau}_a=x^b_a\tau_b,\label{(3.1.41)}
\end{equation}
where $X=(x^b_a)$ is a nondegenerate complex matrix.

Mathematical logic dictates that a physically sound gauge theory
should be invariant under the $SU(N)$ representation transformation
(\ref{(3.1.41)}). Consequently,  the following principle of representation
invariance (PRI) must be universally valid and was first postulated in \cite{field2}.

\begin{pri}[Principle of Representation Invariance]  \la{pr3.6} All
$SU(N)$ gauge theories are invariant under the transformation
(\ref{(3.1.41)}). Namely, the actions of the gauge fields are
invariant and the corresponding gauge field equations as given by
(\ref{(3.1.37)})-(\ref{(3.1.39)}) are covariant under the
transformation (\ref{(3.1.41)}).
\end{pri}

Direct consequences of PRI  include the following; see also \cite{field2} for details:

\begin{itemize}

\item The physical quantities such as $\theta^a,A^a_{\mu}$,
and $\lambda^c_{ab}$ are  $SU(N)$-tensors  under the generator  transformation 
(\ref{(3.1.41)}).

\item  The tensor 
\be \la{gab}
G_{ab}=\frac{1}{4N} \lambda^c_{ad}\lambda^d_{cb} =\frac12 \text{Tr}(\tau_a \tau_b^\dagger)
\ee
is a symmetric positive definite 2nd-order
covariant $SU(N)$-tensor, which can be regarded as a Riemannian metric on $SU(N)$.
 
\item The representation invariant action is 
\begin{align*}
&L=\int_M -\frac14  G_{ab} g^{\mu\alpha}g^{\nu\beta}
F^a_{\mu\nu}F^b_{\alpha\beta}+ \bar{\Psi}\left[ i\gamma^{\mu}(\partial_{\mu}
+ igA^a_{\mu}\tau_a)-m\right] \Psi,
\end{align*}
and the representation invariant gauge field equations are 
\begin{align*}
& G_{ab} \left[ \partial^{\nu}F^b_{\nu\mu}
-  g  \lambda^{b}_{cd} g^{\alpha \beta}F^c_{\alpha\mu}A^d_{\beta}\right]  -  g \bar{\Psi} \gamma_{\mu}\tau_a \Psi =   (\partial_\mu + \alpha_b A^b_\mu) \phi_a, \\
&(i\gamma^{\mu}D_{\mu}- m)\Psi =0.
\end{align*}

 \end{itemize}

As we indicated in \cite{field2},  the field models based on PID appear to be the only model which obeys PRI. In particular, both the standard model and the electroweak theory violate PRI, and consequently they  are approximate models of the fundamental interactions of Nature.

\section{Essentials of the Unified Field Theory Based on PID and PRI}

\subsection{Symmetries}
Symmetry plays a crucial role in physics. In fact, symmetry dictates  and determines 
1) the explicit  form of differential equations governing the underlying physical system,
2)  the space-time  structure of the Universe, and 
 the mechanism of fundamental interactions  of Nature,  and 3)
 conservation laws of the underlying physical system.

Each symmetry is characterized by three main ingredients: space, transformations, and tensors. For gravity, for example, the principle of  general relativity consists of the space-time Riemannian manifold $M$, the general coordinate transformation and the associated tensors. The following are two basic implications of a  symmetry:

\begin{enumerate}

\item[a)]  Fundamental  laws  of Nature  are  universal, and their validity is independent of the space-time location and directions of 
experiments and observations;

\item[b)] The universality of laws of Nature  implies
that the differential equations representing them are covariant. 
Equivalently the Lagrange actions are invariant under  the corresponding 
coordinate transformations.
\end{enumerate}

Laws of the fundamental interactions are dictated by the following 
symmetries:
\be
\begin{aligned}
& \text{gravity:} &&  \text{general relativity},\\
&\text{electromagnetism:}  &&    U(1)\text{ gauge invariance},\\
&\text{weak interaction:} &&      SU(2) \text{ gauge invariance},\\
& \text{strong interaction:} &&   SU(3) \text{ gauge invariance}, 
\end{aligned}\label{(3.1.3)}
\ee

Also,  the last three interactions in (\ref{(3.1.3)}) obey  the
Lorentz invariance and  PRI. As a natural outcome, the three charges $e,
g_w, g_s$   are the coupling constants of $U(1), SU(2), SU(3)$ gauge fields.

Following the simplicity principle of laws of Nature, the three basic  symmetries---the Einstein general relativity, the
Lorentz invariance and the gauge invariance---uniquely determine 
the interaction fields and their Lagrangian actions for the four
interactions, which we describe as follows.

\medskip

\noindent
{\it Gravity}

\medskip

The gravitational fields  are the
Riemannian metric defined on the space-time manifold $M$:
\begin{equation}
ds^2=g_{\mu\nu}dx^{\mu}dx^{\nu},\label{(3.1.4)}
\end{equation}
and $g_{\mu\nu}$ stand for the gravitational potential. The Lagrange
action for the metric (\ref{(3.1.4)}) is the Einstein-Hilbert
functional
\begin{equation}
\mathcal{L}_{EH}=R + \frac{8\pi G}{c^4} S, \label{(3.1.5)}
\end{equation}
where $R$ stands for the scalar curvature of the tangent bundle $TM$  of  $M$.

\medskip

\noindent
{\it Electromagnetism}

\medskip

 The field describing electromagnetic  interaction is the $U(1)$ gauge field
$$A_{\mu}=(A_0,A_1,A_2,A_3),$$
representing the  electromagnetic potential,  and  the Lagrangian  action is
\begin{equation}
\mathcal{L}_{EM}=-\frac{1}{4}A_{\mu\nu}A^{\mu\nu},\label{(3.1.6)}
\end{equation}
which stands for the scalar curvature of the vector bundle 
$M\op \C$. 
Here 
\begin{align*}
& A_{\mu\nu}=\partial_{\mu}A_{\nu}-\partial_{\nu}A_{\mu}.
\end{align*}

\medskip

\noindent
{\it Weak interaction}

\medskip

 The weak fields are the $SU(2)$ gauge fields
$$W^a_{\mu}=(W^a_0,W^a_1,W^a_2,W^a_3)\qfor  1\leq a\leq 3,$$
and their action is
\begin{equation}
\mathcal{L}_W=-\frac{1}{4}G^w_{ab}W^a_{\mu\nu}W^{\mu\nu b},\label{(3.1.7)}
\end{equation}
which also stands for the scalar curvature of spinor bundle:
$M\op (\C^4)^2$. Here
 \begin{align*}
&W^a_{\mu\nu}=\partial_{\mu}W^a_{\nu}-\partial_{\nu}W^a_{\mu}+g_w\lambda^a_{bc}W^b_{\mu}W^c_{\nu} && \text{ for }  1\leq a\leq 3.
\end{align*}

\medskip

\noindent
{\it Strong interaction}

\medskip

 The strong fields are the $SU(3)$ gauge fields
$$S^k_{\mu}=(S^k_0,S^k_1,S^k_2,S^k_3) \qfor 1\leq k\leq 8,$$
and the action is
\begin{equation}
\mathcal{L}_S=-\frac{1}{4}G^s_{kl}S^k_{\mu\nu}S^{\mu\nu l},\label{(3.1.8)}
\end{equation}
which corresponds to the scalar curvature of $M\op (\C^4)^3$.
Here
\begin{align*}
&S^k_{\mu\nu}=\partial_{\mu}S^k_{\nu}-\partial_{\nu}S^k_{\mu}+g_s\Lambda^k_{rl}S^r_{\mu}S^l_{\nu} && \text{ for } 1\leq k\leq 8.
\end{align*}

\subsection{Mechanism of fundamental  interactions}
Albert Einstein was  the first physicist  who postulated  that the gravitational force is caused by the space-time curvature. However,  Yukawa's viewpoint, entirely different from Einstein's, is that the other three fundamental forces take place through exchanging intermediate bosons such as photons for the electromagnetic interaction, W$^\pm$ and Z intermediate vector bosons for the weak interaction, and gluons for the strong interaction. 

Based on the unified field theory presented in this article and in \cite{field2}, in the same spirit as the Einstein's principle of equivalence of gravitational force, it is natural for us to  postulate an alternate mechanism  for all four interactions. 

One crucial component of this viewpoint is  to regard the coupling constants for gauge theories for the electromagnetic, the weak and the strong interactions. In fact, each interaction possesses a charge as follows:
\begin{itemize}

\item The gravitational force   is due to the mass charge  $m$, which is responsible for  all macroscopic motions;

\item The electromagnetic force is due to  the electric charge $e$,
 and holds the atoms and molecules together.

\item The strong force is due to the strong charge $g_s$,  and 
mainly acts on three levels: quarks and gluons, hadrons, and nucleons.

\item The weak force is due to the weak charge $g_w$,  and 
provides the binding energy to hold the mediators, the leptons and quarks
together.
\end{itemize}

Another crucial ingredient for each interaction   is the 
corresponding interaction potential $\Phi$. The relation between
each force $F$,  its associated potential $\Phi$  and  the corresponding charge is given by 
\begin{equation}
F=-g\nabla\Phi ,\label{(3.1.1)}
\end{equation}
where $\nabla$ is the gradient operator in the spatial directions, and $g$ is the interaction charge.


Regarding to the laws of Nature,  physical states are described by  functions
$u=(u_1,\cdots ,u_n)$ defined on the space-time $M$:
\begin{align}
&u:\ M\rightarrow M\op \R^n && \text{ for non-quantum system},\label{(1.1.18)}\\
&u:\ M\rightarrow M\op \C^n  &&  \text{ for quantum system},\label{(1.1.19)}
\end{align}
which are solutions of differential equations associated with the laws of the underlying physical system:
\begin{equation}
\delta L(Du)=0,\label{(1.1.20)}
\end{equation}
where $D$ is a derivative operator,   and $L$ is the Lagrange action. Consider two transformations for  the two physical systems (\ref{(1.1.18)}) and (\ref{(1.1.19)}):
\begin{align}
& \tilde{x}=Lx  && \text{ for\ (\ref{(1.1.18)})},\label{(1.1.21)} \\
& \tilde{u}=e^{i\theta\tau}u &&  \text{ for (\ref{(1.1.19)})},\label{(1.1.22)}
\end{align}
where $x$ is a coordinate system in $M$,  $e^{i\theta\tau}:\ \mathbb{C}^n\rightarrow\mathbb{C}^n$ is
an $SU(n)$ transformation,  and  $\theta$ is a function of $x$, and $\tau$ is a Hermitian matrix.

One important consequence of  the invariance of (\ref{(1.1.20)})
under the transformations (\ref{(1.1.21)}) and (\ref{(1.1.22)})
is  that the derivatives $D$ in (\ref{(1.1.20)})
must take  the  following form:
\begin{align}
& D=\nabla +\Gamma  &&  \text{ for (\ref{(1.1.18)})},\label{(1.1.23)}\\
& D=\nabla +igA && \text{ for\ (\ref{(1.1.19)})},\label{(1.1.24)}
\end{align}
where $\Gamma$ depends on the metrics $g_{ij}$, 
$A$ is a gauge field, representing the interaction potential,
and $g$ is the coupling constant, representing the interaction
charge.

The derivatives defined  in (\ref{(1.1.23)}) and
(\ref{(1.1.24)}) are called connections respectively on $M$ and on the 
complex vector bundle $M\op \C^n$. 
\bt\la{t1.9}
\begin{enumerate}
\item The space $M$ is curved if and only if $\Gamma\neq 0$ in all
coordinates, or equivalently $g_{ij}\neq\delta_{ij}$ under all coordinate systems.

\item The complex bundle $M\op \C^n$ is geometrically nontrivial or
twisted if and only if $A\neq 0$.
\end{enumerate}
\et

Consequently, by Principle of General Relativity, the presence of the gravitational field implies that the space-time manifold is curved, and, by Principle of Gauge Invariance, the presence of the electromagnetic, the weak and strong interactions indicates that the complex vector bundle $M\op \C^n$ is twisted.

This analogy, together with Einstein's vision on gravity as the curved effect of space-time manifold, it is natural for us to postulate the following mechanism for all four interactions. 

\begin{interaction}\la{t1.10}
 The gravitational force is the
curved effect of the time-space, and the electromagnetic, weak,
strong interactions are the twisted effects of the underlying complex vector 
bundles $M\op \C^n$.
\end{interaction}

As mentioned earlier, traditionally one adopts  Yukawa's viewpoint  that 
forces of the interactions of Nature are caused by exchanging 
the field mediators. 

\begin{y-interaction} 
The four fundamental interactions of Nature are mediated by exchanging interaction field particles, called the mediators. The gravitational force is 
mediated by the graviton, the electromagnetic force is  mediated by the
photon, the strong interaction is mediated  by the gluons, and  the weak interaction is mediated by the
intermediate vector bosons $W^{\pm}$ and $Z$.
\end{y-interaction}

It is the Yukawa mechanism that leads to the
$SU(2)$ and $SU(3)$ gauge theories for  the weak and   the strong
interactions. In fact, the three mediators $W^{\pm}$ and $Z$ for
the weak interaction are regarded as the $SU(2)$ gauge fields
$W^a_{\mu}\ (1\leq a\leq 3)$, and the eight gluons for the strong
interaction are considered as the $SU(3)$ gauge fields $S^k_{\mu}\
(1\leq k\leq 8)$. Of course, the three color quantum numbers for the
quarks are an important fact to choose $SU(3)$ gauge theory to
describe the strong interaction.

The two interaction mechanisms  lead to two entirely  different directions
to develop the unified field theory. The need for quantization for all current theories for the four interactions    are based on the
Yukawa Interaction Mechanism. The new unified field theory in this article  is based on the Geometric Mechanism, which focus directly on
the four interaction forces as in (\ref{(3.1.1)}),  and does not involve a 
quantization process.

A radical difference for the two direction mechanisms  is that the Yukawa Mechanism is oriented 
toward to computing  the transition probability for the particle decays
and scatterings, and the Geometric Interaction Mechanism is oriented toward to fundamental laws, such as interaction
potentials, of the four interactions.

\subsection{Geometry of unified fields}
\la{s3.3}

Hereafter  we always assume that the manifold $M$  is the 4-dimensional space-time manifold of our Universe. We adopt the view that symmetry principles determine the geometric structure of $M$,
and the geometries of $M$ associated with the fundamental interactions of Nature  dictate all motion laws defined on $M$. 
The process  that symmetries determine the geometries of $M$ is achieved in
the following three steps:

\begin{enumerate}

\item The symmetric principles, such as the Einstein general relativity, the
Lorentz invariance, and the gauge invariance, determine that the
fields reflecting geometries of $M$ are the Riemannian metric
$\{g_{\mu\nu}\}$ and the gauge fields $\{G^a_{\mu}\}$. In addition,  the
symmetric principles also determine the Lagrangian actions of
$g_{\mu\nu}$ and $G^a_{\mu}$.

\item PID determines the field
equations governing  $g_{\mu\nu}$ and $G^a_{\mu}$.

\item The solutions $g_{\mu\nu}$ and $G^a_{\mu}$ of the field equations
determine the geometries of $M$.
\end{enumerate}

The geometry of unified field   refers to the geometries of $M$, 
determined  by the following known physical symmetry principles:
\be\label{(3.1.18)}
\begin{aligned}
&\text{principle of general relativity}, \\
&\text{principle of Lorentz invariance},\\
&U(1)\times SU(2)\times SU(3) \text{ gauge invariance},  \\
&\text{principle of representation invariance (PRI)}. 
\end{aligned}
\ee

The fields determined by the symmetries in (\ref{(3.1.18)}) are
given by
\begin{itemize}

\item general relativity:  $g_{\mu\nu}:\ M\rightarrow T^0_2M$, the Riemannian metric,

\item Lorentz invariance:   $(\psi ,\Phi ):\ M\rightarrow M\op
[(\C^4)^N\times\C^N]$, the Dirac  and Klein-Golden fields,   

\item $U(1)$ gauge invariance:   $A_{\mu}:\ M\rightarrow T^*M$,  the $U(1)$
   gauge field,

\item $SU(2)$ gauge invariance: $W^a_{\mu}:\ M\rightarrow
(T^*M)^3$,   the $SU(2)$  gauge fields,

\item $SU(3)$ gauge invariance:   $S^k_{\mu}:\ M\rightarrow
(T^*M)^8$, the $SU(3)$   gauge fields. 

\end{itemize}

The Lagrange action for the geometry of the unified fields is given by
\begin{equation}
L=\int_M\left[\mathcal{L}_{EH}+\mathcal{L}_{EM}+\mathcal{L}_W+\mathcal{L}_S+ \mathcal{L}_D+\mathcal{L}_{KG}\right]\sqrt{-g}dx\label{(3.1.20)}
\end{equation}
where  $\mathcal{L}_{EH}$, $\mathcal{L}_{EM}$, $\mathcal{L}_W$  and $\mathcal{L}_S$  are the Lagrangian actions for the four interactions defined by (\ref{(3.1.5)})--(\ref{(3.1.8)}),  and the  actions for both Dirac spinor fields and Klein-Gordon fields $\mathcal{L}_D$  and $\mathcal{L}_{KG}$  are given by 
\be\label{(3.1.21)}
\begin{aligned}
&\mathcal{L}_D=\bar{\Psi}(i\gamma^{\mu}D_{\mu}-m)\Psi,\\
&\mathcal{L}_{KG}=\frac{1}{2}(D^{\mu}\Phi )^{\dag}(D_{\mu}\Phi
)+\frac{1}{2}m^2\Phi^{\dag}\Phi.
\end{aligned}
\ee
Here 
\begin{align*}
&A_{\mu\nu}=\partial_{\mu}A_{\nu}-\partial_{\nu}A_{\mu}, \\
&W^a_{\mu\nu}=\partial_{\mu}W^a_{\nu}-\partial_{\nu}W^a_{\mu}+g_w\lambda^a_{bc}W^b_{\mu}W^c_{\nu} && \text{ for }  1\leq a\leq 3, \\
&S^k_{\mu\nu}=\partial_{\mu}S^k_{\nu}-\partial_{\nu}S^k_{\mu}+g_s\Lambda^k_{rl}S^r_{\mu}S^l_{\nu} && \text{ for } 1\leq k\leq 8,  \\
&\Psi =(\psi^E,\psi^w,\psi^s), \\
&\Phi =(\phi^E,\phi^w,\phi^s), \\
&m=(m_e,m_w,m_s), 
\end{align*}
and
\be\label{(3.1.22)}
\begin{aligned}
&\psi^E:\ M\rightarrow M\op{\C}^4  && \text{1-component Dirac spinor}, \\
&\psi^w:\ M\rightarrow M\op ({\C}^4)^2 &&  \text{2-component Dirac spinors}, \\
&\psi^s:\ M\rightarrow M\op ({\C}^4)^3 && \text{3-component Dirac spinors},\\
&\phi^E:\ M\rightarrow M\op{\C} && \text{1-component Klein-Gordon field}, \\
&\phi^w:\ M\rightarrow M\op{\C}^2 &&  \text{2-component Klein-Gordon fields}, \\
&\phi^s:\ M\rightarrow M\op{\C}^3 && \text{3-component Klein-Gordon fields}.
\end{aligned}
\ee
The derivative operators $D_{\mu}$ are given by 
\begin{eqnarray}
&&D_{\mu}(\psi^E,\phi^E)=(\partial_{\mu}+ieA_{\mu})(\psi^E,\phi^E),\nonumber\\
&&D_{\mu}(\psi^w,\phi^w)=(\partial_{\mu}+ig_wW^a_{\mu}\sigma_a)(\psi^w,\phi^w),\label{(3.1.23)}\\
&&D_{\mu}(\psi^s,\phi^s)=(\partial_{\mu}+ig_sS^k_{\mu}\tau_k)(\psi^s,\phi^s).\nonumber
\end{eqnarray}

The geometry of unified fields consists of 1) the field functions
and 2) the Lagrangian action (\ref{(3.1.20)}), which is invariant under the
following seven transformations:

\begin{enumerate}

\item the general linear transformation on $T_pM$:
\be\label{(3.1.24)}
\begin{aligned}
&\mathcal Q_p = (a^{\mu}_{\nu}):  T_pM\rightarrow T_pM,   && \mathcal Q^{-1}_p=(b^{\mu}_{\nu})^T  \qquad \text{ for }  p\in M, \\
&(\tilde{g}_{\mu\nu})=\mathcal Q_p(g_{\mu\nu})\mathcal Q^T_p, &&\tilde{A}_{\mu}=a^{\nu}_{\mu}A_{\nu},\\
&\tilde{W}^a_{\mu}=a^{\nu}_{\mu}W^a_{\nu}&& \text{for } 1\leq a\leq
3,\\
&\tilde{S}^k=a^{\nu}_{\mu}S^k_{\nu}&& \text{for } 1\leq k\leq
8,  \\
&\tilde{\gamma}^{\mu}=b^{\mu}_{\nu}\gamma^{\nu}, &&
\tilde{\partial}_{\mu}=a^{\nu}_{\mu}\partial_{\nu}, 
\end{aligned}
\ee
and other fields do not change under this transformation. 

\item  Lorentz transformation on $T_pM$:
\be\label{(3.1.25)}
\begin{aligned}
&L= (l^{\nu}_{\mu}):\ T_pM\rightarrow T_pM, &&  L\ \text{is independent\ of}\ p\in
M,  \\
&(\tilde{g}_{\mu\nu})=L(g_{\mu\nu})L^T, 
&&\tilde{A}_{\mu}=l^{\nu}_{\mu}A_{\nu},  \\
&\tilde{W}^a_{\mu}=l^{\nu}_{\mu}W^a_{\nu} && \text{for } 1\leq a\leq 3,\\
&\tilde{S}^k_{\mu}=l^{\nu}_{\mu}S^k_{\nu} && \text{for }  1\leq k\leq 8, \\
&\tilde{\Psi}=R_L\Psi,  && R_L\ \text{is the spinor transformation matrix}, \\
&\tilde{\partial}_{\mu}=l^{\nu}_{\mu}\partial_{\nu},
\end{aligned}
\ee
and other fields do not change under this transformation. 

\item  $U(1)$ gauge transformation on $M\op {\C}^4_p$ and
$M\op {\C}^1_p$:
\be
\begin{aligned}
&\Omega:\ {\C}^4_p\rightarrow{\C}^4_p\   \text{ or } \   \ {\C}^1_p\rightarrow{\C}^1_p && \text{ for }  p\in
M,  \Omega =e^{i\theta}\in U(1),\\
&\tilde{\Psi}^E=e^{i\theta}\psi^E, &&\tilde{\phi}^E=e^{i\theta}\phi^E, \\
&\tilde{A}_{\mu}=A_{\mu}-\frac{1}{e}\partial_{\mu}\theta , 
\end{aligned}\label{(3.1.26)}
\ee

\item  $SU(2)$ gauge transformation:
\be\label{(3.1.27)}
\begin{aligned}
&\Omega :\ ({\C}^4_p)^2\rightarrow
({\C}^4_p)^2 \text{ or }  \ {\C}^2_p\rightarrow \C^2_p,  &&  p\in M,\ \ \ \
\Omega =e^{i\theta^a\sigma_a}\in SU(2), \\
&\tilde{\psi}^w=\Omega\psi^w,  && \tilde{\phi}^w=\Omega\phi^w,\\
&\tilde{W}^a_{\mu}\sigma_a=W^a_{\mu}\Omega\sigma_a\Omega^{-1}+\frac{i}{g_w}\partial_{\mu}\Omega\Omega^{-1}, && \tilde{m}_w=\Omega m_w\Omega^{-1}. 
\end{aligned}\ee

\item  $SU(3)$ gauge transformation:
\be\label{(3.1.28)}
\begin{aligned}
&\Omega :\ ({\C}^4_p)^3\rightarrow
({\C}^4_p)^3, \ \text{ or } \ {\C}^3_p\rightarrow{\C}^3_p,  && \text{ for }  \ p\in
M,\ \ \ \ \Omega =e^{i\theta^k\tau_k}\in SU(3), \\
&\tilde{\psi}^s=\Omega\psi^s,  && \tilde{\phi}^s=\Omega\phi^s, \\
&\tilde{S}^k_{\mu}\tau_k=S^k_{\mu}\Omega\tau_k\Omega^{-1}+\frac{i}{g_s}\partial_{\mu}\Omega\Omega^{-1}, && \tilde{m}_s=\Omega m_s\Omega^{-1}. 
\end{aligned}
\ee

\item $SU(2)$ representation transformation on $T_eSU(2)$:
\be
\begin{aligned}
&X=(x^b_a): T_eSU(2)\rightarrow T_eSU(2), &&  (y^a_b)^T=X^{-1}, \\
&\tilde{\sigma}_s=x^b_a\sigma_b,  &&
(\tilde{G}^w_{ab})=X(G^w_{ab})X^T,   \\
& \tilde{W}^a_{\mu}=y^a_bW^b_{\mu}.\end{aligned}\label{(3.1.29)}
\ee

\item $SU(3)$ representation transformation on $T_eSU(3)$:
\be\label{(3.1.30)}
\begin{aligned}
&X= (x^l_k):\ T_eSU(3)\rightarrow T_eSU(3), && (y^k_l)^T=X^{-1}, \\
&\tilde{\tau}_k=x^l_k\tau_l,  &&(\tilde{G}^s_{kl})=X(G^s_{kl})X^T,\\
&\tilde{S}^k_{\mu}=y^k_lS^l_{\mu}. 
\end{aligned}
\ee

\end{enumerate}


\br\la{r3.3}
Here we adopt the linear transformations of the
bundle spaces instead of the coordinate transformations in the base
manifold $M$. In this case, the two transformations (\ref{(3.1.24)})
and (\ref{(3.1.25)}) are compatible. Otherwise, we have to introduce
the Veibein tensors to overcome the incompatibility  between the
Lorentz transformation and the general coordinate transformation.
\er

\subsection{Gauge symmetry breaking}

In physics, symmetries are displayed in two levels in the laws of Nature:
\begin{align}
&\text{ the\ invariance\ of\ Lagrangian\ actions}\
L,\label{(3.1.32)}\\
&\text{ the\ covariance\ of\ variation\ equations\ of}\
L.\label{(3.1.33)}
\end{align}
The implication of the following three
symmetries:
\be\label{(3.1.34)}
\begin{aligned}
&\text{Einstein\ General\ Relativity}, \\
&\text{Lorentz\ Invariance},\\
&\text{Gauge\ Representation\ Invariance}, 
\end{aligned}
\ee
stands for the universality of physical laws, i.e. the validity of
laws of Nature is independent of the coordinate systems expressing them. Consequently,  the symmetries in (\ref{(3.1.34)}) cannot be broken at 
both levels of (\ref{(3.1.32)}) and (\ref{(3.1.33)}).

However, the physical implication of the gauge symmetry is different at 
the two levels (\ref{(3.1.32)}) and (\ref{(3.1.33)}):

\begin{enumerate}

\item The gauge invariance of the Lagrangian action,  (\ref{(3.1.32)}), amounts to saying that  the  energy contributions of particles in a physical  system are
indistinguishable.

\item The gauge invariance of the variation equations,  (\ref{(3.1.33)}), means that the
particles involved in the  interaction are indistinguishable.

\end{enumerate}

It is clear that the first aspect (1) above is universally true, while the second aspect (2) is not universally  true. In other words,  the Lagrange actions obey
the gauge invariance, but the corresponding  variation equations   break
the gauge symmetry. This suggests us to postulate the following principle of gauge symmetry breaking  for interactions described by a gauge theory.

\begin{pri}[Gauge Symmetry Breaking]\la{pr3.4} The gauge symmetry
holds true only for the Lagrangian actions for the electromagnetic,
week and strong interactions, and it will be violated in the field
equations of these interactions.
\end{pri}

The principle of gauge symmetry breaking can be regarded as part of the spontaneous symmetry breaking, which is a phenomenon
 appearing in various physical fields. In 2008, the Nobel
Prize in Physics was awarded to Y. Nambu for the discovery of the
mechanism of spontaneous symmetry breaking  in subatomic physics. In
2013, F. Englert and P. Higgs were awarded the Nobel Prize for the
theoretical discovery of a mechanism that contributes to our
understanding of the origin of mass of subatomic particles.

Although the phenomenon was discovered in superconductivity by
Ginzburg-Landau in 1951, the mechanism of spontaneous  symmetry breaking  in particle physics was first proposed by Y. Nambu in 1960; see \cite{nambu60, nambu-jona1, nambu-jona2}. The Higgs mechanism, discovered in \cite{higgs, englert, guralnik},  is a special case of the Nambo-Jona-Lasinio spontaneous symmetry breaking,  leading to the mass generation  of sub-atomic particles.

PID discovered in this article provides a new mechanism  for gauge
symmetry breaking and mass  generation. The difference between
both the PID and  the Higgs  mechanisms   is that the first one is a natural
sequence of the first principle, and the second  is to add
artificially a Higgs field in the Lagrangian action. Also, 
the PID mechanism obeys PRI, and the Higgs mechanism violates PRI.  

\section{Unified Field Equations Based on PID and PRI}
\subsection{Unified field equations based on PID}

The abstract unified field equations
(\ref{(3.1.37)})-(\ref{(3.1.38)}) are derived based on PID. 
We now present the detailed form of this model, ensuring that these field 
equations satisfy both the principle of gauge-symmetry breaking and PRI.

By PID, the unified field  model (\ref{(3.1.37)})-(\ref{(3.1.38)})  
are derived as the 
variation of the action (\ref{(3.1.20)}) under the $\text{
div}_A$-constraint
$$
\langle\delta L,X\rangle =0\ \ \ \ \text{ for\ any}\ X\ \text{ with}\
\text{ div}_AX=0.
$$ 
Here it is required that the gradient operator
$\nabla_A$ corresponding to $\text{ div}_A$ are PRI covariant. The gradient operators in different sectors are given as follows:
\be\label{(3.3.6)}
\begin{aligned}
&D^G_{\mu}=\nabla_{\mu}+\alpha^0A_{\mu}+\alpha^1_bW^b_{\mu}+\alpha^2_kS^k_{\mu}, \\
&D^E_{\mu}=\nabla_{\mu}+\beta^0A_{\mu}+\beta^1_bW^b_{\mu}+\beta^2_kS^k_{\mu},\\
&D^w_{\mu}=\nabla_{\mu}+\gamma^0A_{\mu}+\Gamma^1_bW^b_{\mu}+\gamma^2_kS^k_{\mu}-\frac{1}{4}m^2_wx_{\mu}, \\
&D^s_{\mu}=\nabla_{\mu}+\delta^0A_{\mu}+\delta^1_bW^b_{\mu}+\delta^2_kS^k_{\mu}-\frac{1}{4}m^2_sx_{\mu},
\end{aligned}
\ee
where
\begin{equation}
\begin{aligned}
& m_w,m_s,\alpha^0,\beta^0,\gamma^0,\delta^0&&\text{ are scalar
parameters},\\
& \alpha^1_a,\beta^1_a,\gamma^1_a,\delta^1_a &&\text{ are first-order}\
SU(2)\ \text{ tensors},\\
& \alpha^2_k,\beta^2_k,\gamma^2_k,\delta^2_k&&\text{ are first-order}\
SU(3)\ \text{ tensors}.
\end{aligned}\label{(3.3.7)}
\end{equation}
Thus, the PID equations (\ref{(3.1.37)})-(\ref{(3.1.38)}) can be
expressed as
\be
\begin{aligned}
&\frac{\delta L}{\delta
g_{\mu\nu}}=D^G_{\mu}\phi^G_{\nu}, \\
&\frac{\delta L}{\delta A_{\mu}}=D^E_{\mu}\phi^E, \\
&\frac{\delta L}{\delta
W^a_{\mu}}=D^w_{\mu}\phi^w_a,\\
&\frac{\delta L}{\delta S^k_{\mu}}=D^s_{\mu}\phi^s_k, 
\end{aligned}\label{(3.3.8)}
\ee
where $\phi^G_{\nu}$ is a vector field, and $\phi^E, \phi^w,\phi^s$
are scalar fields.

With the PID equations (\ref{(3.3.8)}),  the PRI covariant unified field equations\footnote{We ignore the Klein-Gordon fields.}  are then given as follows:
\begin{align}
&R_{\mu\nu}-\frac{1}{2}g_{\mu\nu}R=-\frac{8\pi
G}{c^4}T_{\mu\nu}+D^G_{\mu}\phi^G_{\nu},\label{(3.3.9)}\\
&\partial^{\mu}(\partial_{\mu}A_{\nu}-\partial_{\nu}A_{\mu})-eJ_{\nu}=D^E_{\nu}\phi^E,\label{(3.3.10)}\\
&G^w_{ab}\left[\partial^{\mu}W^b_{\mu\nu}-g_w\lambda^b_{cd}g^{\alpha\beta}W^c_{\alpha\nu}W^d_{\beta}\right]-g_wJ_{\nu
a}=D^w_{\nu}\phi^w_a,\label{(3.3.11)}\\
&G^s_{kj}\left[\partial^{\mu}S^j_{\mu\nu}-g_s\Lambda^j_{cd}g^{\alpha\beta}S^c_{\alpha\nu}S^d_{\beta}\right]-g_sQ_{\nu
k}=D^s_{\nu}\phi^s_k,\label{(3.3.12)}\\
&(i\gamma^{\mu}D_{\mu}-m)\psi^E=0,\label{(3.3.13)}\\
&(i\gamma^{\mu}D_{\mu}-m_l)\psi^w=0,\label{(3.3.14)}\\
&(i\gamma^{\mu}D_{\mu}-m_g)\psi^s=0,\label{(3.3.15)}
\end{align}
where $D^G_{\mu},D^E_{\nu},D^w_{\nu},D^s_{\nu}$ are given by
(\ref{(3.3.6)}), and
\be\label{(3.3.16)}
\begin{aligned}
&J_{\nu}=\bar{\psi}^E\gamma^{\nu}\psi^E, \\
&J_{\nu
a}=\bar{\psi}^w\gamma^{\nu}\sigma_a\psi^w,\\
&Q_{\nu k}=\bar{\psi}^s\gamma^{\nu}\tau_k\psi^s, \\
&T_{\mu\nu}=\frac{\delta S}{\delta g_{\mu\nu}}+\frac{c^4}{16\pi
G}g^{\alpha\beta}(G^w_{ab}W^a_{\alpha\mu}W^b_{\beta\nu}+G^s_{kl}S^k_{\alpha\mu}S^l_{\beta\nu} +A_{\alpha\mu}A_{\beta\nu}) \\
&\qquad \qquad -\frac{c^4}{16\pi
G}g_{\mu\nu}(\mathcal{L}_{EM}+\mathcal{L}_W+\mathcal{L}_S). 
\end{aligned}
\ee

As mentioned in Section 3, the action
(\ref{(3.1.20)})  for the unified field model is 
invariant under all seven transformations given in Section \ref{s3.3}, including in particular the $U(1)$,  the $SU(2)$, and the $SU(3)$ gauge
transformations.
However, the equations (\ref{(3.3.9)})-(\ref{(3.3.12)}) are not
invariant under the gauge transformations, and spontaneous gauge symmetry breaking is caused by the presence of  the terms
$D^G_{\mu}\phi^G_{\nu}, D^E_{\nu}\phi^E, D^w_{\nu}\phi^w_a,
D^s_{\nu}\phi^s_k$ in the right-hand sides of
(\ref{(3.3.9)})-(\ref{(3.3.12)}) involving  the gauge fields
$A_{\mu}, W^a_{\mu}$ and $S^k_{\mu}$.

In other words,  the unified field model based on PID and PRI obey 
\begin{itemize}
\item principle of general relativity, 
\item the principle of Lorentz invariance, 
\item principle of representation invariance (PRI), 
\item the principle of spontaneous gauge-symmetry breaking. 
\end{itemize}

\subsection{Coupling parameters and physical dimensions}
There are a number of to-be-determined coupling  parameters 
in the  general form of the unified field equations (\ref{(3.3.9)})-(\ref{(3.3.15)}), and the $SU(2)$ and $SU(3)$
generators $\sigma_a$ and $\tau_k$ are taken arbitrarily. 
With PRI we are able to substantially reduce the number of these to-be-determined
parameters in the unified model to two $SU(2)$ and $SU(3)$ tensors
$$
\{\alpha^w_a\}=(\alpha^w_1,\alpha^w_2,\alpha^w_3),\ \ \ \
\{\alpha^s_k\}=(\alpha^s_1,\cdots ,\alpha^s_8),
$$ 
containing 11
parameters, representing the portions distributed to the gauge
potentials by the weak and strong charges.

Also, if we take
$\sigma_a\ (1\leq a\leq 3)$ as the Pauli matrices 
and $\tau_k=\lambda_k\ (1\leq k\leq 8)$ as the Gell-Mann matrices, 
then the two metrics $G^w_{ab}$ and $G^s_{kl}$ are Euclidian:
$$G^w_{ab}=\delta_{ab},  \qquad  G^s_{kl}=\delta_{kl}.$$
Hence, in general we usually take the Pauli matrices $\sigma_a$ and
the Gell-Mann matrices $\lambda_k$ as the $SU(2)$ and $SU(3)$
generators.

For convenience, we first introduce dimensions of related physical
quantities. Let $E$ represent energy, $L$ be the length and $t$ be
the time. Then we have
\begin{align*}
&(A_{\mu},W^a_{\mu},S^k_{\mu}):\ \sqrt{E/L},  &&  (e,g_w,g_s):\
\sqrt{EL},\\
&(J_{\mu},J_{\mu a},Q_{\mu k}):\ 1/L^3, &&
(\phi^E,\phi^w_a,\phi^s_k):\ \frac{\sqrt{E}}{\sqrt{L}L},\\
&(\hbar, c) : (Et, L/t), && {mc}/{\hbar}:\ 1/L.
\end{align*}
In addition, for gravitational fields we have
\begin{equation}
\begin{array}{ll}
g_{\mu\nu}:\ \text{ dimensionless},&R:\ 1/L^2,\\
T_{\mu\nu}:\ E/L^3,&\phi^G_{\mu}:\ 1/L,\\
\text{ gravitational\ constant}&G:\ L^5/Et^4.
\end{array}
\end{equation}
According to the dimensions above, we deduce the dimensions of the 
parameters in (\ref{(3.3.9)})-(\ref{(3.3.15)}) are as follows
\begin{equation}
\begin{aligned}
& (m_w,m_s):1/L,   && \qquad (\alpha^0,\beta^0,\gamma^0,\delta^0):\
1/\sqrt{EL},\\
& (\alpha^1_a,\beta^1_a,\gamma^1_a,\delta^1_a):1/\sqrt{EL},&&\qquad (\alpha^2_k,\beta^2_k,\gamma^2_k,\delta^2_k):\
1/\sqrt{EL}.
\end{aligned}
\end{equation}

Thus the parameters in (\ref{(3.3.7)}) can be rewritten as
\be\label{(3.3.18)}
\begin{aligned}
&(m_w,m_s)=\left(\frac{m_Hc}{\hbar},\frac{m_{\pi}c}{\hbar}\right), \\
&(\alpha^0,\beta^0,\gamma^0,\delta^0)=\frac{e}{\hbar
c}(\alpha^E,\beta^E,\gamma^E,\delta^E), \\
&(\alpha^1_a,\beta^1_a,\gamma^1_a,\delta^1_a)=\frac{g_w}{\hbar
c}(\alpha^w_a,\beta^w_a,\gamma^w_a,\delta^w_a),\\
&(\alpha^2_k,\beta^2_k,\gamma^2_k,\delta^2_k)=\frac{g_s}{\hbar
c}(\alpha^s_k,\beta^s_k,\gamma^s_k,\delta^s_k), 
\end{aligned}
\ee
where $m_H$ and $m_{\pi}$ represent the masses of $\phi^w$ and
$\phi^s$, and all the parameters $(\alpha ,\beta ,\gamma ,\delta )$
on the right hand side of (\ref{(3.3.18)}) with different super and
sub indices are dimensionless constants.

\subsection{Standard form of unified field equations}

Due to (\ref{(3.3.18)}), the unified field equations
(\ref{(3.3.9)})-(\ref{(3.3.15)}) can be simplified in the form
\begin{align}
&R_{\mu\nu}-\frac{1}{2}g_{\mu\nu}R=-\frac{8\pi
G}{c^4}T_{\mu\nu}+\left[\nabla_{\mu}+\frac{e}{\hbar c}\alpha^EA_{\mu} \label{(3.3.19)}   +\frac{g_w}{\hbar
c}\alpha^w_aW^a_{\mu}+\frac{g_s}{\hbar
c}\alpha^s_kS^k_{\mu}\right]\phi^G_{\nu}, \\
&\partial^{\nu}A_{\nu\mu}-eJ_{\nu}=\left[\partial_{\mu}+\frac{e}{\hbar
c}\beta^EA_{\mu}+\frac{g_w}{\hbar
c}\beta^w_aW^a_{\mu}+\frac{g_s}{\hbar
c}\beta^s_kS^k_{\mu}\right]\phi^E,\label{(3.3.20)}\\
&\partial^{\nu}W^a_{\nu\mu}-\frac{g_w}{\hbar
c}\varepsilon^a_{bc}g^{\alpha\beta}W^b_{\alpha\mu}W^c_{\beta}-g_wJ^a_{\mu}\label{(3.3.21)}\\
&\ \ \ \ \ \ \ =\left[\partial_{\mu}+\frac{e}{\hbar
c}\gamma^EA_{\mu}+\frac{g_w}{\hbar
c}\gamma^w_bW^b_{\mu}+\frac{g_s}{\hbar
c}\gamma^s_kS^k_{\mu}-\frac{1}{4}\left(\frac{m_Hc}{\hbar}\right)^2x_{\mu}\right]\phi^a_w,\nonumber\\
&\partial^{\nu}S^k_{\nu\mu}-\frac{g_s}{\hbar
c}f^k_{ij}g^{\alpha\beta}S^i_{\alpha\mu}S^j_{\beta}-g_sQ^k_{\mu}\label{(3.3.22)}\\
&\ \ \ \ \ \ =\left[\partial_{\mu}+\frac{e}{\hbar
c}\delta^EA_{\mu}+\frac{g_w}{\hbar
c}\delta^w_bW^b_{\mu}+\frac{g_s}{\hbar
c}\delta^s_lS^l_{\mu}-\frac{1}{4}\left(\frac{m_{\pi}c}{\hbar}\right)^2x_{\mu}\right]\phi^k_s,\nonumber\\
&(i\gamma^{\mu}D_{\mu}-m)\Psi =0,\label{(3.3.23)}
\end{align}
where $\Psi =(\psi^E,\psi^w,\psi^s)$, and
\be\label{(3.3.24)}
\begin{aligned}
&A_{\mu\nu}=\partial_{\mu}A_{\nu}-\partial_{\nu}A_{\mu}, \\
&W^a_{\mu\nu}=\partial_{\mu}W^a_{\nu}-\partial_{\nu}W^a_{\mu}+\frac{g_w}{\hbar
c}\varepsilon^a_{bc}W^b_{\mu}W^c_{\nu},\\
&S^k_{\mu\nu}=\partial_{\mu}S^k_{\nu}-\partial_{\nu}S^k_{\mu}+\frac{g_s}{\hbar
c}f^k_{ij}S^i_{\mu}S^j_{\nu}. 
\end{aligned}
\ee

Equations (\ref{(3.3.19)})-(\ref{(3.3.23)}) need to be supplemented
with coupled gauge equations to compensate the new dual  fields
$(\phi^E,\phi^a_w,\phi^k_s)$. In different physical situations, the
coupled gauge equations may be different.

From the field theoretical point of view (i.e. not the field particle point of
view), the coefficients in (\ref{(3.3.19)})-(\ref{(3.3.22)}) should
be
\be\label{(3.3.25)}
\begin{aligned}
&(\alpha^w_1,\alpha^w_2,\alpha^w_3)=\alpha^w(\omega_1,\omega_2,\omega_3), \\
&(\beta^w_1,\beta^w_2,\beta^w_3)=\beta^w(\omega_1,\omega_2,\omega_3), \\
&(\gamma^w_1,\gamma^w_2,\gamma^w_3)=\gamma^w(\omega_1,\omega_2,\omega_3), \\
&(\delta^w_1,\delta^w_2,\delta^w_3)=\delta^w(\omega_1,\omega_2,\omega_3),
\end{aligned}
\ee
and
\be\label{(3.3.26)}
\begin{aligned}
&(\alpha^s_1,\cdots ,\alpha^s_8)=\alpha^s(\rho_1,\cdots
,\rho_8), \\
&(\beta^s_1,\cdots ,\beta^s_8)=\beta^s(\rho_1,\cdots
,\rho_8), \\
&(\gamma^s_1,\cdots ,\gamma^s_8)=\gamma^s(\rho_1,\cdots
,\rho_8),\\
&(\delta^s_1,\cdots ,\delta^s_8)=\gamma^s(\rho_1,\cdots
,\rho_8), 
\end{aligned}
\ee
with the unit modules:
\begin{align*}
&|\omega |=\sqrt{\omega^2_1+\omega^2_2+\omega^2_3}=1,\\
&|\rho|=\sqrt{\rho^2_1+\cdots +\rho^2_8}=1,
\end{align*}
using  the Pauli matrices $\sigma_a$ and   the Gell-Mann matrices $\lambda_k$  as the generators for $SU(2)$ and $SU(3)$ respectively.

The two $SU(2)$ and $SU(3)$ tensors in (\ref{(3.3.25)}) and
(\ref{(3.3.26)}),
\begin{equation}
\omega_a=(\omega_1,\omega_2,\omega_3),\ \ \ \ \rho_k=(\rho_1,\cdots
,\rho_8),\label{(3.3.27)}
\end{equation}
are very important, by which we can obtain $SU(2)$ and $SU(3)$
representation invariant gauge fields:
\begin{equation}
W_{\mu}=\omega_aW^a_{\mu},\ \ \ \
S_{\mu}=\rho_kS^k_{\mu}.\label{(3.3.28)}
\end{equation}
which represent respectively the weak and the strong interaction  potentials.

In view of (\ref{(3.3.25)})-(\ref{(3.3.28)}), the unified field
equations for the four fundamental forces are written as
\begin{align}
&R_{\mu\nu}-\frac{1}{2}g_{\mu\nu}R+\frac{8\pi
G}{c^4}T_{\mu\nu}=\left[\nabla_{\mu}+\frac{e\alpha^E}{\hbar
c}A_{\mu}+\frac{g_w\alpha^w}{\hbar
c}W_{\mu}+\frac{g_s\alpha^s}{\hbar
c}S_{\mu}\right]\phi^G_{\nu},\label{(3.3.29)}\\
&\partial^{\nu}A_{\nu\mu}-eJ_{\mu}=\left[\partial_{\mu}+\frac{e\beta^E}{\hbar
c}A_{\mu}+\frac{g_w\beta^w}{\hbar c}W_{\mu}+\frac{g_s\beta^s}{\hbar
c}S_{\mu}\right]\phi^E,\label{(3.3.30)}\\
&\partial^{\nu}W^a_{\nu\mu}-\frac{g_w}{\hbar
c}\varepsilon^a_{bc}g^{\alpha\beta}W^b_{\alpha\mu}W^c_{\beta}-g_wJ^a_{\mu}\label{(3.3.31)}\\
&\ \ \ \ \ \ \
=\left[\partial_{\mu}-\frac{1}{4}k^2_wx_{\mu}+\frac{e\gamma^E}{\hbar
c}A_{\mu}+\frac{g_w\gamma^w}{\hbar
c}W_{\mu}+\frac{g_s\gamma^s}{\hbar
c}S_{\mu}\right]\phi^a_w,\nonumber\\
&\partial^{\nu}S^k_{\nu\mu}-\frac{g_s}{\hbar
c}f^k_{ij}g^{\alpha\beta}S^i_{\alpha\mu}S^j_{\beta}-g_sQ^k_{\mu}\label{(3.3.32)}\\
&\ \ \ \ \ \ \
=\left[\partial_{\mu}-\frac{1}{4}k^2_sx_{\mu}+\frac{e\delta^E}{\hbar
c}A_{\mu}+\frac{g_w\delta^w}{\hbar
c}W_{\mu}+\frac{g_s\delta^s}{\hbar
c}S_{\mu}\right]\phi^k_s,\nonumber\\
&(i\gamma^{\mu}D_{\mu}-m)\Psi =0.\label{(3.3.33)}
\end{align}

\section{Duality and Decoupling of Interaction Fields}

The natural duality of four fundamental interactions to be addressed in this section is
a direct consequence of PID. It is with this 
duality, together with the PRI invariant potentials 
$S_{\mu}$  and  $W_{\mu}$ given by (\ref{(3.5.1)})  and (\ref{w-potential}), that we establish 
a clear  explanation for many longstanding challenging problems in physics, including for example the dark matter and dark energy phenomena, the formulas of the weak and strong forces, the quark confinement, the asymptotic freedom, and the strong
potentials of nucleons. Also, this duality lay a solid foundation for the
weakton model of elementary particles and the energy level theory of
subatomic particles, and give rise to  a new mechanism for sub-atomic decay and scattering.
 
The unified field model can be easily decoupled to study each  individual interaction when other interactions are negligible. In other words, PID is certainly applicable to each individual interaction. For gravity, for example, PID offers  to a new gravitational field model,  leading to a unified model for dark energy and dark matter \cite{MW12}.

\subsection{Duality}
In the unified field equations (\ref{(3.3.19)})-(\ref{(3.3.22)}), 
 there exists a natural duality between the
interaction fields $(g_{\mu\nu},A_{\mu},W^a_{\mu},S^k_{\mu})$ and
their corresponding dual fields $(\phi^G_{\mu},\phi^E,\phi^w_a,\phi^s_k)$ :
\be\label{(3.4.1)}
\begin{aligned} 
& g_{\mu\nu}  &&  \leftrightarrow && \phi^G_{\mu}, \\
&A_{\mu}  &&  \leftrightarrow  &&  \phi^E,\\
&W^a_{\mu} && \leftrightarrow   && \phi^a_w && \text{for } 1\leq a\leq 3, \\
&S^k_{\mu} && \leftrightarrow  &&  \phi^k_s  && \text{for } 1\leq k\leq 8.
\end{aligned}
\ee
Thanks to PRI, the $SU(2)$ gauge fields
$W^a_{\mu}\ (1\leq a\leq 3)$ and the $SU(3)$ gauge fields
$S^k_{\mu}\ (1\leq k\leq 8)$ are symmetric in their indices
$a=1,2,3$ and $k=1,\cdots ,8$ respectively. Therefore, the
corresponding relation (\ref{(3.4.1)}) can be also considered as the
following dual relation
\be\label{(3.4.2)}
\begin{aligned}
& g_{\mu\nu} &&  \leftrightarrow && \phi^G_{\mu},  \\
&A_{\mu}&&  \leftrightarrow && \phi^E,  \\
&\{W^a_{\mu}\}&&  \leftrightarrow &&\{\phi^a_w\},\\
&\{S^k_{\mu}\}&&  \leftrightarrow && \{\phi^k_s\}.
\end{aligned}\ee
The duality  relation (\ref{(3.4.1)}) can be regarded as the
correspondence between field particles for each interaction, and the
relation (\ref{(3.4.2)}) is the duality of interacting forces. We now 
address  these two different dualities.

\medskip

\noindent
{\it Duality of field particles}

\medskip

In the duality relation (\ref{(3.4.1)}),  if the tensor 
fields on the left-hand side are of $k$-th order, then their dual 
tensor fields on the right-hand side are of $(k-1)$-th order. Physically,
this amounts to saying that if a  mediator for an interaction has spin$-k$,
then the dual mediator for the dual  field has spin$-(k-1)$. Hence, (\ref{(3.4.1)}) leads to the following important physical conclusion:

\medskip

\begin{DIM}\la{dim3.13} Each interaction 
mediator possesses  a dual field particle, called the dual mediator,
and if the mediator has spin-$k$, then its dual mediator has spin-$(k-1)$.
\end{DIM}

The duality between interaction  mediators is a direct consequence of PID used for deriving the unified field equations. Based on this duality, if there exist a graviton
with spin $J=2$, then there must exist a dual graviton with spin
$J=1$. In fact, for all interaction mediators,  we have the following duality
correspondence:
\be\label{(3.4.3)}
\begin{aligned}
&\text{graviton}\ (J=2) &&  \leftrightarrow  &&  \text{dual graviton}\ (J=1), \\
&\text{photon}\ (J=1)&&  \leftrightarrow  && \text{dual photon}\ (J=0), \\
&W^{\pm}\ \text{ bosons}\ (J=1)&&  \leftrightarrow  && \text{charged  Higgs}\ H^{\pm}\ (J=0),\\
&Z\ \text{boson}\ (J=1)&&  \leftrightarrow  && \text{neutral\ Higgs}\ H^0\
(J=0), \\
&\text{gluons}\ g^k\ (J=1)&&  \leftrightarrow  && \text{dual gluons}\
\phi^k_g\ (J=0).
\end{aligned}
\ee
 The neutral Higgs $H^0$ (the adjoint particle of $Z$)
had been discovered experimentally. We remark that the duality (\ref{(3.4.3)})
can also be derived using the weakton model \cite{weakton}.

\medskip

\noindent
{\it Duality of interaction forces}

\medskip

The correspondence (\ref{(3.4.2)}) provides a dual relation between
the attracting  and repelling   forces. In fact, from the interaction
potentials we find that the even-spin fields yield attracting  forces,
and the odd-spin fields yield repelling  forces. 

\begin{DIF}\la{t3.14}
Each interaction
generates both attracting and repelling forces. Moreover, for each pair
of dual fields, the even-spin field generates an attracting  force,  and  the
odd-spin field generates a repelling force.
\end{DIF}

This duality of interaction forces is illustrated as follows:
\be
\begin{aligned}
&\text{ Gravitation force} && = && \text{attraction due to }\ g_{\mu\nu} + \text{
repelling  due to }\ \phi^G_{\mu}, \\
&\text{ Electromagnetism} && = && \text{attraction due to }\ \phi^E+\text{ repelling\ due to }\ A_{\mu}, \\
&\text{ Weak\ force} && = && \text{attraction due to }\ \phi_w+\text{ repelling due to }\
W_{\mu},\\
&\text{ Strong\ force} && = && \text{ attraction due to }\ \phi_s+\text{ repelling due to }\
S_{\mu}. 
\end{aligned}\label{(3.4.4)}
\ee

\subsection{Gravitational field equations based on PID}
\la{s5.2}
As we only consider the gravitational interaction, then the
gravitational field equations can be decoupled from the unified
field model (\ref{(3.3.19)})-(\ref{(3.3.23)}), and are given by 
\begin{equation}
R_{\mu\nu}-\frac{1}{2}g_{\mu\nu}R=-\frac{8\pi
G}{c^4}T_{\mu\nu}+(\nabla_{\mu}+\frac{e}{\hbar
c}A_{\mu})\Phi_{\nu},\label{(3.4.5)}
\end{equation}
where the term $\frac{e}{\hbar c}A_{\mu}\Phi_{\nu}$ represents the
coupling between the gravitation and the cosmic microwave background (CMB)
radiation.

Taking  divergence on both sides of (\ref{(3.4.5)}) yields 
\begin{equation}
\nabla^{\mu}\nabla_{\mu}\Phi_{\nu}+\frac{e}{\hbar
c} \nabla^{\mu}(A_{\mu}\Phi_{\nu})=\frac{8\pi
G}{c^4}\nabla^{\mu}T_{\mu\nu}.\label{(3.4.6)}
\end{equation}

The duality of gravitation is based on the  field equations (\ref{(3.4.5)}) and (\ref{(3.4.6)}).

\medskip

\noindent
{\it Gravitons and dual gravitons}

\medskip

It is known that as the equations describing field particles,
(\ref{(3.4.5)}) characterize the graviton as a massless, neutral
bosonic particle with spin $J=2$, and (\ref{(3.4.6)}) indicate that  the dual graviton
is  a massless, neutral bosonic particle with $J=1$. Hence, the gravitational
field equations induced by PID and PRI provide a pair of field
particles:
\begin{equation}
\begin{aligned}
& \text{graviton:} &&J=2,\ m=0,\ Q_e=0,\\
& \text{dual graviton:} &&J=1,\ m=0,\ Q_e=0,
\end{aligned}\label{(3.4.7)}
\end{equation}
where $Q_e$ is the electric charge.

It is the nonlinear interaction of  these two  field particles in (\ref{(3.4.7)}) that lead to the dark matter and dark energy phenomena.

\medskip

\noindent{\it  Gravitational force}

\medskip

We know that from the Schwarzschild solution of the classical Einstein field
equations gives rise to the classical Newton's gravitational force formula:
\begin{equation}
F=-\frac{mMG}{r^2},\label{(3.4.9)}
\end{equation}
which is an attracting force  generated by $g_{\mu\nu}$.

However,  the gravitational force by the field
equations (\ref{(3.4.5)}), then we can deduce a revised formulas to
(\ref{(3.4.9)}). Actually, as ignoring the microwave background
radiation, the equations (\ref{(3.4.5)}) become  \cite{MW12}:
\begin{equation}
R_{\mu\nu}-\frac{1}{2}g_{\mu\nu}R=-\frac{8\pi
G}{c^4}T_{\mu\nu}-\nabla_{\mu}\nabla_{\nu}\varphi ,\label{(3.4.10)}
\end{equation}
where $\Phi_{\nu}=- \nabla_{\nu}\varphi$, and $\varphi$ is a scalar field.
In \cite{MW12}, we are able to derive  from (\ref{(3.4.10)}) that the
gravitational force should be in the form
\begin{equation}
F=mMG\left[-\frac{1}{r^2}+\frac{c^2}{2MG}\Phi
r-\left(\frac{c^2}{MG}+\frac{1}{r}\right)\frac{d\varphi}{dr}\right],\label{(3.4.11)}
\end{equation}
where $\varphi$ is the dual field, representing the scalar potential,
and
\begin{equation}
\Phi =g^{\mu\nu}\nabla_{\mu}\nabla_{\nu}\varphi .\label{(3.4.12)}
\end{equation}
The first term in the right-hand side of (\ref{(3.4.11)}) is the
Newton's gravitational force, and the second term (\ref{(3.4.12)})
represents the repelling force generated by the dual field
$\varphi$, and the third term
$$-\left(\frac{c^2}{MG}+\frac{1}{r}\right)\frac{d\varphi}{dr}$$
represents the force due to the nonlinear coupling of $g_{\mu\nu}$ and its
dual $\varphi$.  Formula (\ref{(3.4.11)}) can be approximatively
written as
\begin{equation}
\begin{aligned}
& F=mMG\left(-\frac{1}{r^2}-\frac{k_0}{r}+k_1r\right),\\
& k_0=4\times 10^{-18}\text{ km}^{-1},\ \ \ \ k_1=10^{-57}\text{ km}^{-3}.
\end{aligned}\label{(3.4.13)}
\end{equation}
The formulas (\ref{(3.4.13)}) shows that a central gravitational
field with mass $M$ has an  attracting force $-k_0/r$ i  addition to the
normal gravitational force, explaining the dark matter, and has a
repelling force $k_1r$, explaining the dark energy; see \cite{MW12} for details.

\subsection{Field equations for strong interactions}

The decoupled field model from (\ref{(3.3.19)})-(\ref{(3.3.23)}) for
strong interactions describing field particles is given by
\begin{align}
&\partial^{\nu}S^k_{\nu\mu}-\frac{g_s}{\hbar
c}f^k_{ij}g^{\alpha\beta}S^i_{\alpha\mu}S^j_{\beta}-g_sQ^k_{\mu}\label{(3.4.33)}\\
&\qquad  =\left[
   \partial_{\mu}+\frac{g_s}{\hbar
c}\delta^s_lS^l_{\mu}-\frac{1}{4}\left(\frac{m_{\pi}c}{\hbar}\right)^2x_{\mu}\right]\phi^k_s \qquad \text{ for } 1\leq k\leq 8,\nonumber\\
&i\gamma^{\mu}\left[ \partial_{\mu}+i\frac{g_s}{\hbar
c}S^k_{\mu}\tau_k \right] \psi -\frac{mc}{\hbar}\psi =0,\label{(3.4.34)}
\end{align}
where $\tau_k=\tau^k$ are the Gell-Mann matrices, and
\begin{equation}
\begin{aligned}
& S^k_{\mu\nu}=\partial_{\mu}S^k_{\nu}-\partial_{\nu}S^k_{\mu}+\frac{g_s}{\hbar
c}S^i_{\mu}S^j_{\nu},\\
& Q^k_{\mu}=\bar{\psi}\gamma^{\mu}\tau^k\psi.
\end{aligned}\label{(3.4.35)}
\end{equation}

Taking divergence on both sides of (\ref{(3.4.33)}) and by
$$\partial^{\mu}\partial^{\nu}S^k_{\mu\nu}=0  \qfor 1\leq k\leq
8,
$$ 
we deduce  the following dual field equations  for the strong interaction:
\begin{align}
&\partial^{\mu}\partial_{\mu}\phi^k_s+\partial^{\mu}\left[\left(\frac{g_s}{\hbar
c}\delta^k_lS^l_{\mu}-\frac{1}{4}\frac{m^2_{\pi}c^2}{\hbar^2}x_{\mu}\right)\phi^k_s\right]\label{(3.4.36)}\\
&\qquad =-g_s\partial^{\mu}Q^k_{\mu}-\frac{g_s}{\hbar
c}f^k_{ij}g^{\alpha\beta}\partial^{\mu}(S^i_{\alpha\mu}S^j_{\beta}).\nonumber
\end{align}

The equations (\ref{(3.4.33)})-(\ref{(3.4.34)}) also need 8 additional gauge 
equations to compensate the induced dual fields $\phi^k_s$:
\begin{equation}
F^k_s(S_{\mu},\phi_s,\psi )=0,\ \ \ \ 1\leq k\leq 8.\label{(3.4.37)}
\end{equation}

We have the following duality for  the strong interaction.

\medskip

\noindent{\it  Gluons and scalar dual gluons}
\medskip

Based on quantum chromodynamics (QCD), the field particles for  the 
strong interaction are eight massless  gluons with spin $J=1$, which
are described by the $SU(3)$ gauge fields $S^k_{\mu}\ (1\leq k\leq
8)$. By the duality (\ref{(3.4.1)}), for the strong
interactions we have the field particle correspondence
$$
S^k_{\mu}\ \leftrightarrow\ \phi^k_s \qfor 1\leq k\leq 8.
$$
It implies that corresponding to the 8 gluons $S^k_{\mu}\ (1\leq
k\leq 8)$ there should be 8 dual gluons represented by $\phi^k_s$,
called the scalar gluons due to $\phi^k_s$ being scalar fields.
Namely we have the following gluon correspondence
$$\text{ gluons}\ g_k\ \leftrightarrow\ \text{ scalar\ gluons}\
\tilde{g}_k\ \ \ \ (1\leq k\leq 8).$$

Gluons and scalar gluons are described by equations (\ref{(3.4.33)})
and (\ref{(3.4.36)}) respectively, which are nonlinear. 
In fact, $g_k$ and $\tilde{g}_k$ are confined in hadrons.

\medskip

\noindent
{\it Strong force}

\medskip

The strong interaction forces are governed by the field equations
(\ref{(3.3.29)})-(\ref{(3.3.33)}). The decoupled field equations are given
by
\begin{align}
&\partial^{\nu}S^k_{\nu\mu}-\frac{g_s}{\hbar
c}f^k_{ij}g^{\alpha\beta}S^i_{\alpha\mu}S^j_{\beta}-g_sQ^k_{\mu}\label{(3.4.38)}
=\left[\partial_{\mu}-\frac{1}{4}k^2_sx_{\mu}+\frac{g_s\delta}{\hbar
c}S_{\mu}\right]\phi^k_s, \\
&\partial^{\mu}\partial_{\mu}\phi^k_s-k^2\phi^k_s+\frac{1}{4}k^2_sx_{\mu}\partial^{\mu}\phi^k_s+\frac{g_s\delta}{\hbar
c}\partial^{\mu}(S_{\mu}\phi^k_s)\label{(3.4.39)}\\
&\qquad  \qquad  =-g_s\partial^{\mu}Q^k_{\mu}-\frac{g_s}{\hbar
c}f^k_{ij}g^{\alpha\beta}\partial^{\mu}(S^i_{\alpha\mu}S^j_{\beta}),\nonumber\\
&i\gamma^{\mu}\left[ \partial_{\mu}+i\frac{g_s}{\hbar
c}S^l_{\mu}\tau_l\right] \psi -\frac{mc}{\hbar}\psi =0,\label{(3.4.40)}
\end{align}
for $1 \le k \le 8$, where $\delta$ is a parameter. 

\br\la{r3.16}
Usually, $k_s$ and $\delta$  are 
regarded as masses of the field particles. However, since 
(\ref{(3.4.38)})-(\ref{(3.4.40)}) are the field equations for the
interaction forces,  the parameters $k_s$ and $\delta$ are 
no longer viewed  masses. In fact, $k^{-1}$ represents the
range of attracting force for the strong interaction, and
$\left(\frac{g_s\phi^0_s}{\hbar c}\delta \right)^{-1}$ is the
range of the repelling force, where $\phi^0_s$ is a ground state of
$\phi_s$.
\er

Thanks to PRI, the strong interaction potential takes the following linear
combination of the eight $SU(3)$ gauge fields:
\begin{equation}
S_{\mu}=\rho_kS^k_{\mu},\label{(3.5.1)}
\end{equation}
where $\rho_k=(\rho_1,\cdots ,\rho_8)$ is the $SU(3)$ tensor as
given in (\ref{(3.3.27)}).

Let $g_s$ be the strong charge of an elementary particle, equivalent to the strong charge of $w^*$ weakton  as introduced in \cite{weakton}, and  let 
$$
\Phi_0=S_0\ \text{ the  temporal-component  of (\ref{(3.5.1)})}
$$
be the strong charge potential of this particle. Then the strong
force between two elementary particles with strong charges is
$$F=-g_s\nabla\Phi_0.$$

However, the strong interactions are layered, i.e. the strong forces
act only on particles at  the same level, such as quarks and quarks, hadrons
and hadrons, etc. Hence, the strong interaction potentials are also
layered. In fact, we have derived in \cite{strong}  the  layered formulas of strong interaction potentials. In particular,  the
$w^*$-weakton potential $\Phi_0$, the quark potential $\Phi_q$, the
nucleon/hadron potential $\Phi_n$ and the atom/molecule potential
$\Phi_a$  are given as follows \cite{strong}:
\be\label{(3.5.41)}
\begin{aligned}
&\Phi_0=g_s\left[\frac{1}{r}-\frac{A_0}{\rho_w}(1+k_0r)e^{-k_0r}\right], \\
&\Phi_q=\left(\frac{\rho_w}{\rho_q}\right)^3g_s\left[\frac{1}{r}-\frac{A_q}{\rho_q}(1+k_1r)e^{-k_1r}\right],\\
&\Phi_n=3\left(\frac{\rho_w}{\rho_n}\right)^3g_s\left[\frac{1}{r}-\frac{A_n}{\rho_n}(1+k_nr)e^{-k_nr}\right], \\
&\Phi_a=N\left(\frac{\rho_w}{\rho_a}\right)^3g_s\left[\frac{1}{r}-\frac{A_a}{\rho_a}(1+k_ar)e^{-k_ar}\right]. 
\end{aligned}
\ee
Here, $k_0,k_1,k_n,k_a$ are  given by  
\be
\begin{aligned}
& \frac{1}{k_0}=10^{-18}\text{ cm},  &&  \qquad \frac{1}{k_1}=10^{-16}\text{
cm}, \\
&  \frac{1}{k_n}=10^{-13}\text{ cm}, && \qquad 
\frac{1}{k_a}=10^{-10}\sim 10^{-7}\text{ cm}.
\end{aligned}\label{(3.5.42)}
\ee

\subsection{Weak interaction field equations}
Unified field model can be decoupled to study the weak interaction only, leading to the following weak interaction  field equations:
\begin{align}
&\partial^{\nu}W^a_{\nu\mu}-\frac{g_w}{\hbar
c}\varepsilon^a_{bc}g^{\alpha\beta}W^b_{\alpha\mu}W^c_{\beta}-g_wJ^a_{\mu}\label{(3.4.41)}\\
&\qquad \qquad 
=\left[\partial_{\mu}-\frac{1}{4}\left(\frac{m_Hc}{\hbar}\right)^2x_{\mu}+\frac{g_w}{\hbar
c}\gamma^w_bW^b_{\mu}\right] \phi^a_w,\nonumber\\
&i\gamma^{\mu}\left[ \partial_{\mu}+i\frac{g_w}{\hbar
c}W^a_{\mu}\sigma_a\right] \psi -\frac{mc}{\hbar}\psi =0,\label{(3.4.42)}
\end{align}
where $m_H$ represents  the mass of the Higgs particle, 
$\sigma_a=\sigma^a\ (1\leq a\leq
3)$ are the Pauli matrices 
and
\begin{equation}
\begin{aligned}
& W^a_{\mu\nu}=\partial_{\mu}W^a_{\nu}-\partial_{\nu}W^a_{\mu}+\frac{g_w}{\hbar
c}\varepsilon^a_{bc}W^b_{\mu}W^c_{\nu},\\
& J^a_{\mu}=\bar{\psi}\gamma^{\mu}\sigma^a\psi.
\end{aligned}\label{(3.4.43)}
\end{equation}

Taking divergence on both sides of (\ref{(3.4.41)}) we get
\begin{eqnarray}
&&\partial^{\mu}\partial_{\mu}\phi^a_w-\left(\frac{m_Hc}{\hbar}\right)^2\phi^a_w+\frac{g_w}{\hbar
c}\gamma^w_b\partial^{\mu}(W^b_{\mu}\phi^a_w)-\frac{1}{4}(\frac{m_Hc}{\hbar})^2x_{\mu}\partial^{\mu}\phi^a_w\label{(3.4.44)}\\
&&\ \ \ \ =-\frac{g_w}{\hbar
c}\varepsilon^a_{bc}g^{\alpha\beta}\partial^{\mu}(W^b_{\alpha\mu}W^c_{\beta})-g_w\partial^{\mu}J^a_{\mu}.\nonumber
\end{eqnarray}

Also, we need to supplement   (\ref{(3.4.41)})-(\ref{(3.4.42)}) with three additional 3
gauge equations:
\begin{equation}
F^a_w(W_{\mu},\phi_w,\psi )=0\qfor 1\leq a\leq 3.\label{(3.4.45)}
\end{equation}

\bigskip

\noindent{\it Duality between $W^{\pm}, Z$ Bosons and Higgs Bosons $H^{\pm}, H^0$}

\medskip

The three massive vector bosons, denoted by $W^{\pm}, Z^0$, has been discovered experimentally. The field equations (\ref{(3.4.41)}) give rise to a natural duality:\begin{equation}
Z^0\ \leftrightarrow\ H^0,\ \ \ \ W^{\pm}\ \leftrightarrow\
H^{\pm}, \label{(3.4.46)}
\end{equation}
where $H^0, H^{\pm}$  are three dual scalar bosons,  called the
Higgs particles. The neutral Higgs $H^0$ has been discovered by LHC in 2012, and the charged Higgs $H^{\pm}$  have yet to found experimentally.

\medskip

\noindent{\it Weak force}

\medskip

If consider the weak interaction force, we have to use the
 equations decoupled from (\ref{(3.3.29)})-(\ref{(3.3.33)}):
\begin{align}
&\partial^{\nu}W^a_{\nu\mu}-\frac{g_w}{\hbar
c}\varepsilon^a_{bc}g^{\alpha\beta}W^b_{\alpha\mu}W^c_{\beta}-g_wJ^a_{\mu} =  \left[ \partial_{\mu}-\frac{1}{4}k^2_wx_{\mu}+\frac{g_w}{\hbar
c}\gamma W_{\mu}\right] \phi^a_w, \label{(3.4.47)} \\
&\partial^{\mu}\partial_{\mu}\phi^a_w-k^2\phi^2_w+\frac{g_w}{\hbar
c}\gamma\partial^{\mu}(W_{\mu}\phi^a_w)-\frac{1}{4}k^2x_{\mu}\partial^{\mu}\phi^a_w\label{(3.4.48)}\\
&\ \ \ \ =-g_w\partial^{\mu}J^a_{\mu}-\frac{g_w}{\hbar
c}\varepsilon^a_{bc}g^{\alpha\beta}\partial^{\mu}(W^b_{\alpha\mu}W^c_{\beta}),\nonumber\\
&i\gamma^{\mu}(\partial_{\mu}+i\frac{g_w}{\hbar
c}W^a_{\mu}\sigma_a)\psi -\frac{mc}{\hbar}\psi =0,\label{(3.4.49)}
\end{align}
where $\gamma ,k_w$ are constants. 

As in the case for  the strong interaction,  the weak
interaction potential is given by the following PRI  invariant
\be\la{w-potential}
W_{\mu}=\omega_aW^a_{\mu}=(W_0,W_1,W_2,W_3),
\ee
where $\omega_a\ (1\leq a\leq 3)$ is the $SU(2)$ tensor as in
(\ref{(3.3.27)}).
The weak charge potential and weak force are as
\begin{equation}
\begin{aligned}
&  \Phi_w=W_0   &&\text{ the\ time\ component\ of}\ W_{\mu},\\
& F_w=-g_w(\rho )\nabla\Phi_w,
\end{aligned}\label{(3.6.1)}
\end{equation}
where $g_w(\rho )$ is the weak charge of a particle with radius
$\rho$.

We have derived  \cite{weak} from  (\ref{(3.4.47)})-(\ref{(3.4.49)})  the following layered weak interaction potential formulas:
\begin{equation}
\begin{aligned}
& \Phi_w=g_w(\rho)e^{-kr}\left[\frac{1}{r}-\frac{B}{\rho}(1+2kr)e^{-kr}\right],\\
& g_w(\rho )=N\left(\frac{\rho_w}{\rho}\right)^3g_w,
\end{aligned}\label{(3.6.16)}
\end{equation}
where $\Phi_w$ is the weak force potential of a particle with radius
$\rho$ and $N$ weak charges $g_w$, $g_w$ is the unit weak charge of weak charge
$g_w$ for each weakton \cite{weakton},  $\rho_w$ is the
weakton radius, $B$ is a parameter depending on the particles, and
\begin{equation}
\frac{1}{k}=10^{-16}\text{ cm},\label{(3.6.17)}
\end{equation}
represents the force-range of the weak interaction.

\section{Orthogonal Decomposition for Tensor Fields}
\la{s2.3}

\subsection{Orthogonal decomposition theorems}
The aim of this  section is to derive an orthogonal decomposition
for $(k,r)$-tensor fields, with $k+r\geq 1$, into divergence-free and
gradient parts. This decomposition plays a crucial role for the
unified field theory introduced in this paper.

Let $M$ be a closed Riemannian manifold or $M=S^1\times\tilde{M}$ be
a closed Minkowski manifold with metric 
\begin{equation}
(g_{\mu\nu})=\left(\begin{matrix} -1&0\\
0&G\end{matrix}\right). \label{(2.3.44)}
\end{equation}
Here $\tilde M$ is  a closed 
Riemannian manifold, 
and $G=(g_{ij})$ is the Riemannian metric of $\tilde{M}$.

Let $A$ be a vector field or a covector field, and $u\in
L^2(T^k_rM)$. We define the operators $D_A$ and ${\rm div}_A$ by 
\begin{equation}
\begin{aligned}
& D_Au=Du+u\otimes A,\\
& {\rm div}_Au={\rm div}u-u\cdot A.
\end{aligned}\label{(2.2.39)}
\end{equation}

A tensor field  $v\in L^2(T^k_rM)\ (k+r\geq 1)$ is $\text{
div}_A$-free, denoted by $\text{ div}_Av=0$, if
\begin{equation}
\int_M(\nabla_A\psi ,v)\sqrt{g}dx=0\qquad \text{ for any } \psi \text{ with } \nabla_A\psi\in
L^2(T^k_rM). \label{(2.3.9)}
\end{equation}
Here $\psi\in H^1(T^{k-1}_rM)$ or $H^1(T^k_{r-1}M), \nabla_A$ and
$\text{ div}_A$ are as in (\ref{(2.2.39)}).

We remark that if $v\in H^1(T^k_rM)$ satisfies (\ref{(2.3.9)}), then
$v$ is weakly differentiable, and $\text{ div}\ v=0$ in $L^2$-sense.
If $v\in L^2(T^k_rM)$ is not differentiable, then (\ref{(2.3.9)})
means that $v$ is $\text{ div}_A$-free in the distribution sense.

\bt[Orthogonal Decomposition Theorem]\la{t2.17}
Let $A$ be
a given vector field or covector field, and $u\in L^2(T^k_rM)$. Then
the following assertions hold true:

\begin{enumerate}

\item The tensor field $u$ can be orthogonally decomposed into
\begin{equation}
u=\nabla_A\varphi +v\ \ \ \ \text{ with}\ \text{
div}_Av=0,\label{(2.3.10)}
\end{equation}
where $\varphi\in H^1(T^{k-1}_rM)$ or $\varphi\in H^1(T^k_{r-1}M)$.

\item If $M$ is a compact Riemannian manifold, then $u$ can be
orthogonally decomposed into
\begin{equation}
u=\nabla_A\varphi +v+h,\label{(2.3.11)}
\end{equation}
where $\varphi$ and $v$ are as in (\ref{(2.3.10)}), and $h$ is a
harmonic field, i.e.
$$\text{ div}_Ah=0,\ \ \ \ \nabla_Ah=0.$$
In particular, the subspace of all harmonic tensor fields in
$L^2(T^k_rM)$ is of finite dimensional:
\begin{equation}
\begin{aligned}
& H(T^k_rM)=\{h\in L^2(T^k_rM)|\ \nabla_Ah=0,\ \text{ div}_Ah=0\},\ \text{
and}\\
& \text{ div}\ H(T^k_rM)<\infty .
\end{aligned}\label{(2.3.12)}
\end{equation}
\end{enumerate}
\et

\br\la{r2.18}
The above orthogonal decomposition theorem
implies that $L^2(T^k_rM)$ can be decomposed into
\begin{equation}
\begin{aligned}
& L^2(T^k_rM)=G(T^k_rM)\oplus L^2_D(T^k_rM)\ \ \ \ \text{ for\ general\
case},\\
& L^2(T^k_rM)=G(T^k_rM)\oplus H(T^k_rM)\oplus L^2_N(T^k_rM)\ \ \ \
\text{ for}\ M\ \text{ compact\ Riemainnian}.
\end{aligned}\label{(2.3.13)}
\end{equation}
Here $H$ is as in (\ref{(2.3.12)}), and
\begin{align*}
&G(T^k_rM)=\{v\in L^2(T^k_rM)|\ v=\nabla_A\varphi ,\ \varphi\in
H^1(T^k_{r-1}M)\},\\
&L^2_D(T^k_rM)=\{v\in L^2(T^k_rM)|\ \text{ div}_Av=0\},\\
&L^2_N(T^k_rM)=\{v\in L^2_D(T^k_rM)|\ \nabla_Av\neq 0\}.
\end{align*}
They are orthogonal to each other:
$$L^2_D(T^k_rM)\bot G(T^k_rM),\ \ \ \ L^2_N(T^k_rM)\bot H(T^k_rM),\
\ \ \ G(T^k_rM)\bot H(T^k_rM).$$
\er

\br\la{r2.19}
The orthogonal decomposition (\ref{(2.3.13)})
of $L^2(T^k_rM)$ implies that if a tensor field $u\in L^2(T^k_rM)$
satisfies that
$$\langle u,v\rangle_{L^2}=\int_M(u,v)\sqrt{g}dx=0 \qquad 
\forall\text{ div}_Av=0,
$$ 
then $u$ must be a gradient field, i.e.
$$u=\nabla_A\varphi\ \ \ \ \text{ for\ some}\ \varphi\in
H^1(T^{k-1}_rM)\ \text{ or}\ H^1(T^k_{r-1}M).
$$ 
Likewise, if $u\in
L^2(T^k_rM)$ satisfies that
$$\langle u,v\rangle_{L^2}=0 \qquad  \forall v\in G(T^k_rM),$$
then $u\in L^2_D(T^k_rM)$. It is the reason why we define a $\text{
div}_A$-free field by (\ref{(2.3.9)}).
\er

\bp[Proof of Theorem \ref{t2.17}]
 We proceed in several steps as
follows.

\medskip

{\sc Step 1. Proof of Assertion (1).}  Let $u\in L^2(E), E=T^k_rM$ $(k+r\geq
1)$. Consider the equation
\begin{equation}
\Delta\varphi =\text{ div}_Au\ \ \ \ \text{ in}\ M,\label{(2.3.14)}
\end{equation}
where $\Delta$ is the Laplace operator defined by
\begin{equation}
\Delta =\text{ div}_A\cdot\nabla_A.\label{(2.3.15)}
\end{equation}

Without loss of generality, we only consider the case where $\text{
div}_Au\in\tilde{E}=T^{k-1}_rM$. It is clear that if  
(\ref{(2.3.14)}) has a solution $\varphi\in H^1(\tilde{E})$, then by
(\ref{(2.3.15)}), the following vector field must be $\text{
div}_A$-free
\begin{equation}
v=u-\nabla_A\varphi\in L^2(E).\label{(2.3.16)}
\end{equation}
Moreover, by (\ref{(2.3.9)}), we have
\begin{equation}
\langle v,\nabla_A\psi\rangle_{L^2}=\int_M(v,\nabla_A\psi
)\sqrt{g}dx=0.\label{(2.3.17)}
\end{equation}
Namely $v$ and $\nabla_A\varphi$ are orthogonal. Therefore, the
orthogonal decomposition $u=v+\nabla_A\varphi$ follows from
(\ref{(2.3.16)}) and (\ref{(2.3.17)}).

It suffices then to prove that (\ref{(2.3.14)}) has a weak solution
$\varphi\in H^1(\tilde{E})$:
\begin{equation}
\langle\nabla_A\varphi -u,\nabla_A\psi\rangle_{L^2}=0\qquad 
\forall\psi\in H^1(\tilde{E}).\label{(2.3.18)}
\end{equation}

Obviously, if $\phi$ satisfies
\begin{equation}
\Delta\phi =0,\label{(2.3.19)}
\end{equation}
where   $\Delta$  is as  in (\ref{(2.3.15)}),
then, by   integration by parts, 
$$\int_M(\Delta\phi ,\phi )\sqrt{g}dx=-\int_M(\nabla_A\phi
,\nabla_A\phi )\sqrt{g}dx=0.$$ 
Hence (\ref{(2.3.19)}) is equivalent
to
\begin{equation}
\nabla_A\phi =0.\label{(2.3.20)}
\end{equation}
Therefore, for all $\phi$ satisfying (\ref{(2.3.19)}) we have
$$\int_M(u,\nabla_A\phi )\sqrt{g}dx=0.$$
By the Fredholm alternative theorem, 
we derive that the equation (\ref{(2.3.14)}) has a
unique weak solution $\phi\in H^1(\tilde{E})$. 

For Minkowski manifolds,  the existence of solutions for  (\ref{(2.3.14)}) is classical.  Assertion (1) is proved.

\medskip

{\sc  Step 2. Proof of Assertion (2).} 
Based on Assertion (1), we have
\begin{align*}
&H^k(E)=H^k_D\oplus G^k, && L^2(E)=L^2_D\oplus G,
\end{align*}
where
\begin{align*}
&H^k_D=\{u\in H^k(E)|\ \text{ div}_Au=0\},\\
&G^k=\{u\in H^k(E)|\ u=\nabla_A\psi\}. 
\end{align*} 
Define an
operator $\tilde{\Delta}:\ H^2_D(E)\rightarrow L^2_D(E)$ by
\begin{equation}
\tilde{\Delta} u=P\Delta u,\label{(2.3.21)}
\end{equation}
where $P:\ L^2(E)\rightarrow L^2_D(E)$ is the canonical orthogonal
projection.

We known that the Laplace operator $\Delta$ can be expressed as
\begin{equation}
\Delta =\text{ div}_A\cdot\nabla_A=g^{kl}\frac{\partial^2}{\partial
x^k\partial x^l}+B,\label{(2.3.22)}
\end{equation}
where $B$ is a lower-order differential  operator. Since $M$ is
compact, the Sobolev embeddings
$$H^2(E)\hookrightarrow H^1(E)\hookrightarrow L^2(E)$$
are compact. Hence  the lower-order differential  operator
$$B:\ H^2(M,\R^N)\rightarrow L^2(M,\R^N) $$
is a linear compact operator. Therefore the operator in
(\ref{(2.3.22)}) is a linear completely continuous field
$$\Delta :\ H^2(E)\rightarrow L^2(E),$$
which implies that the operator of (\ref{(2.3.21)}) is also a linear
completely continuous field
$$\tilde{\Delta}=P\Delta :\ H^2_D(E)\rightarrow L^2_D(E).$$
By the spectrum theorem of completely continuous  fields \cite{b-book},
the space
$$\tilde{H}=\{u\in H^2_D(E)|\ \tilde{\Delta}u=0\}$$
is finite dimensional, and is the eigenspace of the eigenvalue
$\lambda =0$. By integration by parts, for $u\in\tilde{H}$ we have
\begin{align*}
\int_M(\tilde{\Delta}u,u)\sqrt{g}dx
=&\int_M(\Delta u,u)\sqrt{g}dx\
\ \ \ (\text{ by\ div}_Au=0)\\
=&-\int_M(\nabla_Au,\nabla_Au)\sqrt{g}dx\\
=&0\ \ \ \ (\text{ by}\ \tilde{\Delta}u=0).
\end{align*}
It follows that
$$u\in\tilde{H}\qquad  \Leftrightarrow\qquad  \nabla_Au=0,$$
which implies that $\tilde{H}$ is the same as the harmonic space $H$
of (\ref{(2.3.12)}), i.e. $\tilde{H}=H$. Thus we have
\begin{align*}
&L^2_D(E)=H\oplus L^2_N(E),\\
&L^2_N(E)=\{u\in L^2_D(E)|\ \nabla_Au\neq 0\}.
\end{align*}
The proof of Theorem \ref{t2.17} is complete.
\ep

\subsection{Orthogonal decomposition on manifolds with boundary}

In the above subsections, we mainly consider the orthogonal
decomposition of tensor fields on the closed Riemannian and
Minkowski manifolds. In this subsection we discuss the problem on
manifolds with boundary.

\medskip

\noindent
{\it  Orthogonal decomposition on Riemannian manifolds with boundaries}

\medskip

\bt\la{t2.21} Let $M$ be a Riemannian manifold with boundary $\partial
M\neq\emptyset$, and
\begin{equation}
u:\ M\rightarrow T^k_rM\label{(2.3.37)}
\end{equation}
be a $(k,r)$-tensor field.
Then  we have the following orthogonal decomposition:
\begin{align}
&u=\nabla_A\varphi +v,\label{(2.3.38)}\\
&\text{ div}_Av=0,\qquad  v\cdot n|_{\partial M}=0,\qquad 
\int_M(\nabla_A\varphi ,v)\sqrt{g}dx=0, \nonumber 
\end{align}
where $\partial v/\partial n=\nabla_Av\cdot n$ is the derivative of
$v$ in the direction of outward normal vector $n$ on
$\partial\Omega$.
\et

\bp For the tensor field $u$ in (\ref{(2.3.37)}), consider 
\begin{equation}
\begin{aligned}
& \text{ div}_A\cdot\nabla_A\varphi =\text{ div}_Au && \forall x\in M,\\
& \frac{\partial\varphi}{\partial n}=u\cdot n && \forall x\in\partial M.
\end{aligned}\label{(2.3.39)}
\end{equation}
This Neumann boundary problem possesses  a
solution provided the following condition holds true:
\begin{equation}
\int_{\partial M}\frac{\partial\varphi}{\partial n}ds=\int_{\partial
M}u\cdot nds, \label{(2.3.40)}
\end{equation}
which  is ensured by the boundary
condition in (\ref{(2.3.39)}). Hence by (\ref{(2.3.39)}) the field
\begin{equation}
v=u-\nabla_A\varphi\label{(2.3.41)}
\end{equation}
is $\text{ div}_A$-free, and satisfies the boundary condition
\begin{equation}
v\cdot n|_{\partial M}=0.\label{(2.3.42)}
\end{equation}
Then it follows from (\ref{(2.3.41)}) and (\ref{(2.3.42)}) that the
tensor field $u$ in (\ref{(2.3.37)}) can be orthogonally decomposed
into the form of (\ref{(2.3.38)}). The proof is complete.
\ep

\medskip

\noindent
{\it Orthogonal decomposition on Minkowski manifolds}

\medskip

Let $M$ be a Minkowski  manifold in the form
\begin{equation}
M=\tilde{M}\times (0,T),\label{(2.3.43)}
\end{equation}
with the metric (\ref{(2.3.44)})

In view of the Minkowski metric (\ref{(2.3.44)}), we see that
the operator $\text{ div}_A\cdot\nabla_A$ is a hyperbolic differential
operator expressed as
\begin{equation}
\text{ div}_A\cdot\nabla_A=-\left(\frac{\partial}{\partial
t}+A_0\right)^2+g^{ij}D_{Ai}D_{Aj}.\label{(2.3.45)}
\end{equation}

Now a tensor field $u\in L^2(T^k_rM)$ has an orthogonal composition
if the following hyperbolic equation
\begin{equation}
\text{ div}_A\cdot\nabla_A\varphi =\text{ div}_Au\ \ \ \ \text{ in}\
M\label{(2.3.46)}
\end{equation}
has a weak solution $\varphi\in H^1(T^{k-1}_rM)$ in the following sense:
\begin{equation}
\int_M(D_A\varphi ,D_A\psi )\sqrt{-g}dx=\int_M(u,D_A\psi
)\sqrt{-g}dx \qquad  \forall 
\psi\in H^1(T^{k-1}_rM).
\label{(2.3.47)}
\end{equation}

\bt\la{t2.22}
Let $M$ be a Minkowskian manifold as defined
by (\ref{(2.3.43)})-(\ref{(2.3.44)}), and $u\in L^2(T^l_rM) $ $(k+r\geq
1)$ be an $(k,r)-$tensor field. Then $u$ can be orthogonally
decomposed into the following form
\begin{equation}
\begin{aligned}
& u=\nabla_A\varphi +v,\ \ \ \ \text{ div}_Av=0,\\
& \int_M(\nabla_A\varphi ,v)\sqrt{-g}dx=0,
\end{aligned}\label{(2.3.48)}
\end{equation}
if and only if  equation (\ref{(2.3.46)}) has a weak solution
$\varphi\in H^1(T^{k-1}M)$  in the sense of 
(\ref{(2.3.47)}).
\et

\section{Variations with $\text{ div}_A$-Free Constraints}
Let $M$ be a closed manifold. A Riemannian metric $G$ on $M$ is a
mapping
$$G:\ M\rightarrow T^0_2M=T^*M\otimes T^*M,$$
which is symmetric and nondegenerate. Namely, in a local coordinate
system, $G$ can be expressed as
\begin{equation}
G=\{g_{ij}\}\ \ \ \ \text{ with}\ \ \ \ \
g_{ij}=g_{ji},\label{(2.4.27)}
\end{equation}
and the matrix $(g_{ij})$ is invertible on $M$:
$$G^{-1} =(g^{ij})=(g_{ij})^{-1}: M\to  T^2_0M=TM\otimes TM.$$

If we regard a Riemannian metric $G=\{g_{ij}\}$ as a tensor field on
the manifold $M$, then the set of all metrics $G=\{g_{ij}\}$ on $M$
constitute a topological space, called the space of Riemannian
metrics on $M$. 
The space of Riemannian metrics on $M$ is defined by
\begin{align*}
W^{m,2}(M,g) =& \{G|\ G\in W^{m,2}(T^0_2M),\ G^{-1}\in
W^{m,2}(T^2_0M),\\
 &G\ \text{ is\ the\ Riemannian\ metric\ on}\ M\ \text{ as\ in\
(\ref{(2.4.27)})}\}.
\end{align*}
The space $W^{m,2}(M,g)$ is a metric space, but not a Banach space.
However, it is a subspace of the direct sum of two Sobolev spaces
$W^{m,2}(T^0_2M)$ and $W^{m,2}(T^2_0M)$:
$$W^{m,2}(M,g)\subset W^{m,2}(T^0_2M)\oplus W^{m,2}(T^2_0M).$$

A functional defined on $W^{m,2}(M,g):$
\begin{equation}
F:\ W^{m,2}(M,g)\rightarrow \R\label{(2.4.28)}
\end{equation}
is called the functional of Riemannian metric. In general, the
functional (\ref{(2.4.28)}) can be expressed as
\begin{equation}
F(g_{ij})=\int_Mf(g_{ij},\cdots
,\partial^mg_{ij})\sqrt{g}dx.\label{(2.4.29)}
\end{equation}
Since $(g^{ij})$ is the inverse of $(g_{ij})$, we have
\begin{equation}
g_{ij}=\frac{1}{g}\times\text{ the\ cofactor\ of}\
g^{ij}.\label{(2.4.30)}
\end{equation}
Therefore, $F(g_{ij})$ in (\ref{(2.4.29)}) also depends on $g^{ij}$,
i.e. putting (\ref{(2.4.30)}) in (\ref{(2.4.29)}) we get
\begin{equation}
F(g^{ij})=\int_M\tilde{f}(g^{ij},\cdots
,\partial^mg^{ij})\sqrt{g}dx.\label{(2.4.31)}
\end{equation}

We note that although $W^{m,2}(M,g)$ is not a linear space, for a
given element $g_{ij}\in W^{m,2}(M,g)$ and any symmetric tensor
fields $X_{ij},X^{ij}$, there is a number $\lambda_0>0$ such that
\begin{equation}
\begin{aligned}
& g_{ij}+\lambda X_{ij}\in W^{m,2}(M,g)  && \forall 0\leq |\lambda
|<\lambda_0,\\
& g^{ij}+\lambda X^{ij}\in W^{m,2}(M,g)  &&  \forall 0\leq |\lambda
|<\lambda_0.
\end{aligned}\label{(2.4.32)}
\end{equation}

With (\ref{(2.4.32)}), we can define the  following derivative operators of
the functional $F$:
\begin{align*}
&\delta_*F:\ W^{m,2}(M,g)\rightarrow W^{-m,2}(T^2_0M),\\
&\delta^*F:\ W^{m,2}(M,g)\rightarrow W^{-m,2}(T^0_2M),
\end{align*}
where $W^{-m,2}(E)$ is the dual space of $W^{m,2}(E)$, and
$\delta_*F, \delta^*F$ are defined by
\begin{align}
&\langle\delta_*F(g_{ij}), X\rangle
=\eld F(g_{ij}+\lambda X_{ij}),\label{(2.4.33)}\\
&\langle\delta^*F(g^{ij}),X\rangle
=\eld F(g^{ij}+\lambda X^{ij}).\label{(2.4.34)}
\end{align}

For any give metric $g_{ij}\in W^{m,2}(M,g)$, the value of
$\delta_*F$ and $\delta^*F$ at $g_{ij}$ are second-order
contra-variant and covariant tensor fields:
\begin{equation}
\begin{aligned}
& \delta_*F(g_{ij}):\ M\rightarrow TM\times TM,\\
& \delta^*F(g_{ij}):\ M\rightarrow T^*M\times T^*M.
\end{aligned}\label{(2.4.35)}
\end{equation}

\bt\la{t2.23}
Let $F$ be the functionals defined by
(\ref{(2.4.29)}) and (\ref{(2.4.31)}). Then the following assertions
hold true:

\begin{enumerate}

\item For any $g_{ij}\in W^{m,2}(M,g), \delta_*F(g_{ij})$ and
$\delta^*F(g_{ij})$ are symmetric tensor fields.

\item  If $\{g_{ij}\}\in W^{m,2}(M,g)$ is an extremum point of $F$, i.e.
$\delta F(g_{ij})=0$, then $\{g^{ij}\}$ is also an extremum point of
$F$.

\item $\delta_*f$ and $\delta^*F$ have the following relation
$$(\delta^*F(g_{ij}))^{kl}=-g^{kr}g^{ls}(\delta_*F(g_{ij}))_{rs},$$
where $(\delta^*F)^{kl}$ and $(\delta_*F)_{kl}$ are the components
of $\delta^*F$ and $\delta_*F$.
\end{enumerate}
\et

\bp
We only need to verify Assertion (3).  In view of 
$g_{ik}g^{kj}=\delta^j_i,$
we have the variational relation
$$\delta (g_{ik}g^{kj})=g_{ik}\delta g^{kj}+g^{kj}\delta g_{ik}=0.$$
It implies that
\begin{equation}
\delta g^{kl}=-g^{ki}g^{lj}\delta g_{ij}.\label{(2.4.36)}
\end{equation}
In addition, in (\ref{(2.4.33)}) and (\ref{(2.4.34)}),
$$\lambda X_{ij}=\delta g_{ij},\ \ \ \ \lambda X^{ij}=\delta
g^{ij},\ \ \ \ \lambda\neq 0\ \text{ small}.
$$ 
Therefore, by
(\ref{(2.4.36)}) we get
\begin{align*}
\langle (\delta_*F)_{kl},\delta g^{kl}\rangle
=&-\langle
(\delta_*F)_{kl},g^{ki}g^{lj}\delta g_{ij}\rangle\\
=&-\langle g^{ki}g^{lj}(\delta_*F)_{kl},\delta g_{ij}\rangle\\
=&\langle (\delta^*F)^{ij},\delta g_{ij}\rangle .
\end{align*}
Hence we have
$$(\delta^*F)^{ij}=-g^{ki}g^{lj}(\delta_*F)_{kl}.$$
Thus Assertion (3) follows and the proof is complete.
\ep

We are now in  position to consider the variation with $\text{
div}_A$-free constraints. We know that an extremum point $g_{ij}$ of
a metric functional is a solution of the equation
\begin{equation}
\delta F(g_{ij})=0,\label{(2.4.37)}
\end{equation}
in the  sense  that 
\begin{align}
\langle\delta
F(g_{ij}),X\rangle
=&\eld F(g_{ij}+\lambda
X_{ij})|_{\lambda =0}\nonumber\\
=&\int_M(\delta^*F(g_{ij}))^{kl}X_{kl}\sqrt{g}dx\nonumber\\
=&0 \qquad \qquad  \forall X_{kl}=X_{lk}\in L^2(T^0_2M).\label{(2.4.38)}
\end{align}
Note that the solution $g_{ij}$ of (\ref{(2.4.37)}) in the usual sense
should satisfy
\begin{equation}
\langle\delta F(g_{ij}),X\rangle =0\qquad  \forall X\in
L^2(T^0_2M).\label{(2.4.39)}
\end{equation}
Notice that  (\ref{(2.4.38)}) has
a symmetric constraint  on the variational elements $X_{ij}$:
$X_{ij}=X_{ji}$.
Therefore, comparing  (\ref{(2.4.38)}) with (\ref{(2.4.39)}),  we
may wonder if a  solution $g_{ij}$ satisfying (\ref{(2.4.38)}) is also  a solution of (\ref{(2.4.38)}).  Fortunately,  note that $L^2(T^0_2M)$ can be
decomposed into a direct sum of symmetric and anti-symmetric
spaces as follows
\begin{align*}
&L^2(T^0_2M)=L^2_s(T^0_2M)\oplus L^2_c(T^0_2M),\\
&L^2_s(T^0_2M)=\{u\in L^2(T^0_2M)|\ u_{ij}=u_{ji}\},\\
&L^2_c(T^0_2M)=\{u\in L^2(T^0_2M)|\ u_{ij}=-u_{ji}\},
\end{align*}
and $L^2_s(T^0_2M)$ and $L^2_c(T^0_2M)$ are orthogonal:
\begin{align*}
\int_Mg^{ik}g^{jl}u_{ij}v_{kl}\sqrt{g}dx=&-\int_Mg^{ik}g^{jl}u_{ij}v_{lk}\sqrt{g}dx\\
=&0 \ \ \ \ \forall u\in L^2_s(T^0_2M),\ \ \ \ v\in L^2_c(T^0_2M).
\end{align*}
Thus, due to the symmetry of $\delta F(g_{ij})$, the solution
$g_{ij}$ of (\ref{(2.4.37)}) satisfying (\ref{(2.4.38)}) must also
satisfy (\ref{(2.4.39)}). Hence the solutions of (\ref{(2.4.37)}) in
the sense of (\ref{(2.4.38)}) are the solutions in the usual sense.

However, if we consider the variations of $F$ under the $\text{
div}_A$-free constraint, then the extremum points of $F$ are not
solutions of (\ref{(2.4.37)}) in the usual sense. 
Motived by physical considerations, 
we now introduce   variations with $\text{ div}_A$-free
constraints.

\bd\la{d2.24}
Let $F(u)$ be a functional of tensor fields
$u$. We say that $u_0$ is an extremum point of $F(u)$ under the
$\text{ dev}_A$-free constraint, if
\begin{equation}
\langle\delta F(u_0),X\rangle =\eld F(u_0+\lambda
X)=0 \qquad \forall\text{ div}_AX=0,\label{(2.4.40)}
\end{equation}
where $\text{ div}_A$ is as defined in (\ref{(2.2.39)}). 

In
particular, if $F$ is a functional of Riemannian metrics, and the
solution $u_0=g_{ij}$ is a Riemannian metric, then the differential
operator $D_A$ in $\text{ div}_AX$ in (\ref{(2.4.40)}) is given by
\begin{equation}
D_A=D+A,\ \ \ \ D=\partial +\Gamma ,\label{(2.4.41)}
\end{equation}
and the connection $\Gamma$ is taken at the extremum point
$u_0=g_{ij}$.
\ed

We have the following theorems for $\text{ div}_A$-free
constraint variations.

\bt\la{t2.25}
Let $F:\ W^{m,2}(M,g)\rightarrow \R^1$ be a
functional of Riemannian metrics. Then there is a vector field
$\Phi\in H^1(TM)$ such that the extremum points $\{g_{ij}\}$ of $F$ with
the $\text{ div}_A$-free constraint   satisfy  the
equation
\begin{equation}
\delta F(g_{ij})=D\Phi +A\otimes\Phi, \label{(2.4.42)}
\end{equation}
where $D$ is the covariant derivative operator as in
(\ref{(2.4.41)}).
\et

\bt\la{t2.26}
 Let $F:\ H^m(TM)\rightarrow \R^1$ be a
functional of vector fields. Then there is a scalar  function
$\varphi\in H^1(M)$ such that for a given vector field $A$, the
extremum points $u$ of $F$ with the $\text{ div}_A$-free constraint
satisfy the equation
\begin{equation}
\delta F(u)=(\partial +A)\varphi .\label{(2.4.43)}
\end{equation}
\et

\bp[Proof of Theorems \ref{t2.25} and \ref{t2.26}]  First we prove Theorem \ref{t2.25}.
By (\ref{(2.4.40)}), the extremum points $\{g_{ij}\}$ of $F$ with the 
$\text{ div}_A$-free constraint satisfy
$$\int_M\delta F(g_{ij})\cdot X\sqrt{g}dx=0\qquad 
 \forall X\in
L^2(T^2_0M)\ \text{ with}\ \text{ div}_AX=0.$$ 
It implies that
\begin{equation}
\delta F(g_{ij})\bot L^2_D(T^0_2M)=\{v\in L^2(T^0_2M)|\ \text{
div}_Av=0\}.\label{(2.4.44)}
\end{equation}
By Theorem \ref{t2.17}, $L^2(T^0_2M)$ can be orthogonally decomposed into
\begin{align*}
&L^2(T^0_2M)=L^2_D(T^0_2M)\oplus G^2(T^0_2M),\\
&G^2(T^0_2M)=\{D_A\Phi |\ \Phi\in H^1(T^0_1M)\}.
\end{align*}
Hence it follows from (\ref{(2.4.44)}) that
$$\delta F(g_{ij})\in G^2(T^0_2M),$$
which means that the equality (\ref{(2.4.42)}) holds true.

To prove Theorem \ref{t2.26}, for an 
extremum vector field $u$ of $F$ with the $\text{ div}_A$-free
constraint, we derive in the same fashion that $u$ satisfies the following equation
\begin{equation}
\int_M\delta F(u)\cdot X\sqrt{g}dx=0\qquad  \forall X\in L^2(TM)\
\text{ with}\ \text{ div}_Ax=0.\label{(2.4.45)}
\end{equation}
In addition, Theorem \ref{t2.17} means that
\begin{align*}
&L^2(TM)=L^2_D(TM)\oplus G^2(TM),\\
&L^2_D(TM)=\{v\in L^2(TM)|\ \text{ div}_Av=0\},\\
&G^2(TM)=\{D_A\varphi |\ \varphi\in H^1(M)\}.
\end{align*}
Then we infer from (\ref{(2.4.45)}) that
$$\delta F(u)\in G^2(TM).$$
Thus we deduce the equality (\ref{(2.4.43)}).

The proofs of Theorems \ref{t2.25} and \ref{t2.26} are complete.
\ep

\bibliographystyle{siam}

\begin{thebibliography}{10}

\bibitem{englert}
{\sc F.~Englert and R.~Brout}, {\em Broken symmetry and the mass of gauge
  vector mesons}, Physical Review Letters, 13 (9) (1964), p.~321Ð23.

\bibitem{griffiths}
{\sc D.~Griffiths}, {\em Introduction to elementary particles}, Wiley-Vch,
  2008.

\bibitem{guralnik}
{\sc G.~Guralnik, C.~R. Hagen, and T.~W.~B. Kibble}, {\em Global conservation
  laws and massless particles}, Physical Review Letters, 13 (20) (1964),
  p.~585Ð587.

\bibitem{halzen}
{\sc F.~Halzen and A.~D. Martin}, {\em Quarks and leptons: an introductory
  course in modern particle physics}, John Wiley and Sons, New York, NY, 1984.

\bibitem{higgs}
{\sc P.~W. Higgs}, {\em Broken symmetries and the masses of gauge bosons},
  Physical Review Letters, 13 (1964), p.~508Ð509.

\bibitem{kaku}
{\sc M.~Kaku}, {\em Quantum Field Theory, A Modern Introduction}, Oxford
  University Press, 1993.

\bibitem{kane}
{\sc G.~Kane}, {\em Modern elementary particle physics}, vol.~2, Addison-Wesley
  Reading, 1987.

\bibitem{b-book}
{\sc T.~Ma and S.~Wang}, {\em Bifurcation theory and applications}, vol.~53 of
  World Scientific Series on Nonlinear Science. Series A: Monographs and
  Treatises, World Scientific Publishing Co. Pte. Ltd., Hackensack, NJ, 2005.

\bibitem{strong}
\leavevmode\vrule height 2pt depth -1.6pt width 23pt, {\em Duality theory of
  strong interaction}, Indiana University Institute for Scientific Computing
  and Applied Mathematics Preprint Series, \#1301:
  \url{http://www.indiana.edu/~iscam/preprint/1301.pdf},  (2012).

\bibitem{weak}
\leavevmode\vrule height 2pt depth -1.6pt width 23pt, {\em Duality theory of
  weak interaction}, Indiana University Institute for Scientific Computing and
  Applied Mathematics Preprint Series, \#1302:
  \url{http://www.indiana.edu/~iscam/preprint/1302.pdf},  (2012).

\bibitem{field2}
\leavevmode\vrule height 2pt depth -1.6pt width 23pt, {\em Unified field theory
  and principle of representation invariance}, arXiv:1212.4893; version 1
  appeared in Applied Mathematics and Optimization, DOI:
  10.1007/s00245-013-9226-0, 33pp.,  (2012).

\bibitem{MW12}
\leavevmode\vrule height 2pt depth -1.6pt width 23pt, {\em Gravitational field
  equations and theory of dark matter and dark energy}, Discrete and Continuous
  Dynamical Systems, Ser. A, 34:2 (2014), pp.~335--366; see also
  arXiv:1206.5078v2.

\bibitem{weakton}
\leavevmode\vrule height 2pt depth -1.6pt width 23pt, {\em Weakton model of
  elementary particles and decay mechanisms}, Indiana University Institute for
  Scientific Computing and Applied Mathematics Preprint Series, \#1304:
  \url{http://www.indiana.edu/~iscam/preprint/1304.pdf},  (May 30, 2013).

\bibitem{nambu60}
{\sc Y.~Nambu}, {\em Quasi-particles and gauge invariance in the theory of
  superconductivity}, Phys. Rev., 117 (1960), pp.~648--663.

\bibitem{nambu-jona1}
{\sc Y.~Nambu and G.~Jona-Lasinio}, {\em Dynamical model of elementary
  particles based on an analogy with superconductivity. {I}}, Phys. Rev., 122
  (1961), pp.~345--358.

\bibitem{nambu-jona2}
\leavevmode\vrule height 2pt depth -1.6pt width 23pt, {\em Dynamical model of
  elementary particles based on an analogy with superconductivity. {II}}, Phys.
  Rev., 124 (1961), pp.~246--254.

\bibitem{quigg}
{\sc C.~Quigg}, {\em Gauge theories of the strong, weak, and electromagnetic
  interactions, 2nd edition}, Princeton Unversity Press, 2013.

\end{thebibliography}

\end{document}